\documentclass[12pt]{article}

\usepackage{graphicx,amsmath,amsfonts,amssymb,dcolumn,amsthm,slashed,bbm,xspace,pgffor,tikz,mathbbol,latexsym}
\usepackage[colorlinks,citecolor=blue,urlcolor=blue,linkcolor=blue]{hyperref}

\usepackage[hang]{footmisc} 

\numberwithin{equation}{section}

\begin{document}

\title{\textbf{Topological Yang-Mills theories in self-dual and anti-self-dual Landau gauges revisited}}

\author{\textbf{O.~C.~Junqueira$^1$}\thanks{octavio@if.uff.br}\ , \textbf{A.~D.~Pereira$^1$}\thanks{aduarte@if.uff.br}\ , \textbf{G.~Sadovski$^{1,2}$}\thanks{gsadovski@if.uff.br}\  , \\ \textbf{R.~F.~Sobreiro$^1$}\thanks{sobreiro@if.uff.br}\ , \textbf{A.~A.~Tomaz$^{1,3}$}\thanks{tomaz@cbpf.br}\\\\
\textit{{\small $^1$UFF - Universidade Federal Fluminense, Instituto de F\'isica,}}\\
\textit{{\small Campus da Praia Vermelha, Av. General Milton Tavares de Souza s/n, 24210-346,}}\\
\textit{{\small Niter\'oi, RJ, Brasil.}}\\
\textit{{\small $^2$CECs - Centro de Estudios Cient\'ificos}}\\
\textit{{\small Arturo Prat 514, Valdivia, Chile.}}\\
\textit{{\small $^3$CBPF - Centro Brasileiro de Pesquisas F\'isicas,}}\\
\textit{{\small Rua Dr. Xavier Sigaud 150, 22290-180,}}\\
\textit{{\small Rio de Janeiro, RJ, Brasil.}}}
\date{}
\maketitle
 
\begin{abstract}
We reconsider the renormalizability of topological Yang-Mills field theories in (anti-)self-dual Landau gauges. By employing algebraic renormalization techniques we show that there is only one independent renormalization. Moreover, due to the rich set of Ward identities, we are able to obtain some important exact features of the (connected and one-particle irreducible) two-point functions. Specifically, we show that all two-point functions are tree-level exact.
\end{abstract}

\newpage
\tableofcontents
\newpage

\section{Introduction}

Topological Yang-Mills theories are, essentially, a pure gauge fixing term, \emph{i.e.}, an exact BRST term. In four Euclidean dimensions, such actions are usually known as Donaldson-Witten models \cite{Donaldson:1983wm,Witten:1988ze,Birmingham:1991ty}. As discussed, for instance, in \cite{Baulieu:1988xs,Brooks:1988jm,Myers:1989ur}, topological Yang-Mills theories require three different gauge fixings, two for the gauge field and one for the topological ghost. Typically, one fixes the divergence of the gauge field, the field strength and the divergence of the topological ghost. The renormalization properties of such models were investigated in \cite{Brooks:1988jm,Birmingham:1988ap,WerneckdeOliveira:1993pa,Brandhuber:1994uf}. Particularly in \cite{Brandhuber:1994uf}, the renormalizability at all orders in perturbation theory was proven for (anti-)self-dual Landau gauges. By (anti-)self-dual Landau gauges we mean that the gauge field and the topological ghost are both transverse and that the field strength is equal to ($\pm$) its dual. Remarkably, the authors found only four independent renormalizations. Moreover, they also found that the theory has a topological vector supersymmetry which, together with BRST symmetry and translations, form a supersymmetry algebra of Wess-Zumino type.  

In the present work we study the renormalizability of (anti-)self-dual Landau gauges, --  the same gauge considered in \cite{Brandhuber:1994uf}. For that, we employ the algebraic renormalization techniques \cite{Piguet:1995er}. As a first novelty, we consider two non-trivial extra symmetries that are present in this class of gauges. The main consequence of these extra symmetries is that the model actually has only one independent renormalization. Further consequences of the rich set of Ward identities are explored in terms of the two-point connected Green functions and the one-particle irreducible (1PI) functions. We are able to show directly from the Ward identities that many of these functions are tree-level exact. In particular, the topological gluon (gauge field) two-point functions vanish to all orders. 

Concerning the Faddeev-Popov ghost two-point functions, it is shown that they are equal to the bosonic ghost (ghost of the ghost) two-point functions. In the same way, the topological ghost two-point functions are equal to the mixed two-point functions of the gauge field and a Lagrange multiplier (the one which implements the gauge-fixing for field strength). Nevertheless, from an explicit loop expansion, we are able to show that all of these two-point functions are also tree-level exact. 

This paper is organized as follows: in Section 2, we discuss the gauge fixing structure of the classical Witten's model and describe its quantization; in Section 3, we provide the construction of the most general counterterm by taking into account the rich set of Ward identities and provide the proof of quantum stability; Section 4 is devoted to the analysis of the two-point functions and their properties to all orders in perturbation theory. The special cases of the Faddeev-Popov and topological ghosts two-point functions as well as the explicit computation of the tree-level gauge field propagator are discussed in Section 5. Finally, Section 6 contains our concluding remarks.  

\section{Topological Yang-Mills theories and gauge fixing}

Following \cite{Baulieu:1988xs}, a topological action $S_o[A]$ is invariant under (infinitesimal) gauge transformations of the form
\begin{equation}
\delta A^a_\mu=D_\mu^{ab}\alpha^b+\alpha_\mu^a\;,\label{gt1}
\end{equation}
where $A_\mu^a$ is the gauge field (topological gluon), $D_\mu^{ab}=\delta^{ab}\partial_\mu-gf^{abc}A^c_\mu$ is the covariant derivative in the adjoint representation of the gauge group, $g$ is the coupling constant, $\alpha^a$ and $\alpha_\mu^a$ are the gauge parameters. The four-dimensional spacetime is assumed to be Euclidean and flat. The parameter $\alpha^a$ is associated to a semi-simple Lie group $G$ while $\alpha_\mu^a$ is also $G$-valued and characterizes all other possible transformations associated to the fact that $S_o[A]$ is a topological invariant. A direct consequence of the transformation law \eqref{gt1} is that the field strength,
\begin{equation}
F_{\mu\nu}^a=\partial_\mu A^a_\nu-\partial_\nu A^a_\mu+gf^{abc}A^b_\mu A^c_\nu\;,\label{fs1}
\end{equation}
also transforms as a gauge field,
\begin{equation}
\delta F^a_{\mu\nu}=-gf^{abc}\alpha^bF_{\mu\nu}^c+D_\mu^{ab}\alpha_\nu^b-D_\nu^{ab}\alpha_\mu^b\;.\label{gt2}
\end{equation}
It will be important in what follows to define the dual field strength,
\begin{equation}
\widetilde{F}_{\mu\nu}^a=\frac{1}{2}\epsilon_{\mu\nu\alpha\beta}F_{\alpha\beta}^a\;.\label{fs2}
\end{equation}
Moreover, an ambiguity is present in the gauge parameter $\alpha_\mu^a$, which is defined modulo an extra gauge transformation
\begin{equation}
\delta \alpha_\mu^a=D_\mu^{ab}\lambda^b\;.\label{gt3}
\end{equation}

The model can be consistently quantized through the BRST quantization method. For that, the gauge parameters are promoted to ghost fields: $\alpha^a\longrightarrow c^a$, $\alpha_\mu^a\longrightarrow\psi^a_\mu$, and $\lambda^a\longrightarrow\phi^a$. The field $c^a$ is recognized as the usual Faddeev-Popov ghost, $\psi^a_\mu$ is the topological ghost and $\phi^a$ is the ghost of the ghost, or simply, bosonic ghost. The corresponding BRST transformations are given by
\begin{eqnarray}
sA_\mu^a&=&-D_\mu^{ab}c^b+\psi^a_\mu\;,\nonumber\\
sc^a&=&\frac{g}{2}f^{abc}c^bc^c+\phi^a\;,\nonumber\\
s\psi_\mu^a&=&gf^{abc}c^b\psi^c_\mu+D_\mu^{ab}\phi^b\;,\nonumber\\
s\phi^a&=&gf^{abc}c^b\phi^c\;.\label{brst1}
\end{eqnarray}
As it can easily be seen from \eqref{gt1}, \eqref{gt2} and \eqref{gt3}, there are three gauge symmetries to be fixed. Following \cite{Brandhuber:1994uf}, we employ the (anti-)self-dual Landau gauges,
\begin{eqnarray}
\partial_\mu A_\mu^a&=&0\;,\nonumber\\
\partial_\mu\psi^a_\mu&=&0\;,\nonumber\\
F^a_{\mu\nu}\pm\widetilde{F}_{\mu\nu}^a&=&0\;.\label{gf1}
\end{eqnarray}
Hence, we need three BRST doublets to enforce the three gauge fixings, namely,
\begin{eqnarray}
s\bar{c}^a&=&b^a\;,\;\;\;\;\;\;\;\;sb^a\;=\;0\;,\nonumber\\
s\bar{\chi}^a_{\mu\nu}&=&B_{\mu\nu}^a\;,\;\;sB_{\mu\nu}^a\;=\;0\;,\nonumber\\
s\bar{\phi}^a&=&\bar{\eta}^a\;,\;\;\;\;\;\;\;s\bar{\eta}^a\;=\;0\;,\label{brst2}
\end{eqnarray}
where $\bar{\chi}^a_{\mu\nu}$ and $B_{\mu\nu}^a$ are (anti-)self-dual fields\footnote{The relation depends on the sign on the gauge fixing of the field strength in \eqref{gf1}. For instance, if the sign is positive in \eqref{gf1}, the doublet fields $(\bar{\chi},B)$ are self-dual.}. For completeness and further use, the quantum numbers of all fields are displayed in Table \ref{table1}.

\begin{table}[h]
\centering
\setlength{\extrarowheight}{.5ex}
\begin{tabular}{cc@{\hspace{-.3em}}cccccccccc}
\cline{1-1} \cline{3-12}
\multicolumn{1}{|c|}{Field} & & \multicolumn{1}{|c|}{$A$} & \multicolumn{1}{c|}{$\psi$} & \multicolumn{1}{c|}{$c$} & \multicolumn{1}{c|}{$\phi$} & \multicolumn{1}{c|}{$\bar{c}$} & \multicolumn{1}{c|}{$b$} &\multicolumn{1}{c|}{$\bar{\phi}$} & \multicolumn{1}{c|}{$\bar{\eta}$} & \multicolumn{1}{c|}{$\bar{\chi}$} & \multicolumn{1}{c|}{$B$}   
\\ \cline{1-1} \cline{3-12} 
\\[-1.18em]
\cline{1-1} \cline{3-12}
\multicolumn{1}{|c|}{Dim} & & \multicolumn{1}{|c|}{1} & \multicolumn{1}{c|}{1} & \multicolumn{1}{c|}{0} & \multicolumn{1}{c|}{0} & \multicolumn{1}{c|}{2} & \multicolumn{1}{c|}{2} &\multicolumn{1}{c|}{2} & \multicolumn{1}{c|}{2} & \multicolumn{1}{c|}{2} & \multicolumn{1}{c|}{2}   
\\
\multicolumn{1}{|c|}{Ghost n$^o$} & & \multicolumn{1}{|c|}{0} & \multicolumn{1}{c|}{1} & \multicolumn{1}{c|}{1} & \multicolumn{1}{c|}{2} & \multicolumn{1}{c|}{-1} & \multicolumn{1}{c|}{0} &\multicolumn{1}{c|}{-2} & \multicolumn{1}{c|}{-1} & \multicolumn{1}{c|}{-1} & \multicolumn{1}{c|}{0}   
\\ \cline{1-1} \cline{3-12} 
\end{tabular}
\caption{Quantum numbers of the fields.}
\label{table1}
\end{table}

The complete gauge fixing action is then given by
\begin{eqnarray}
S_{gf}&=&s\int d^4z\left[\bar{c}^a\partial_\mu A_\mu^a+\frac{1}{2}\bar{\chi}^a_{\mu\nu}\left(F_{\mu\nu}^a\pm\widetilde{F}_{\mu\nu}^a\right)+\bar{\phi}^a\partial_\mu\psi^a_\mu\right]\nonumber\\
&=&\int d^4z\left[b^a\partial_\mu A_\mu^a+\frac{1}{2}B^a_{\mu\nu}\left(F_{\mu\nu}^a\pm\widetilde{F}_{\mu\nu}^a\right)+
\left(\bar{\eta}^a-\bar{c}^a\right)\partial_\mu\psi^a_\mu+\bar{c}^a\partial_\mu D_\mu^{ab}c^b+\right.\nonumber\\
&-&\left.\frac{1}{2}gf^{abc}\bar{\chi}^a_{\mu\nu}c^b\left(F_{\mu\nu}^c\pm\widetilde{F}_{\mu\nu}^c\right)-\bar{\chi}^a_{\mu\nu}\left(\delta_{\mu\alpha}\delta_{\nu\beta}\pm\frac{1}{2}\epsilon_{\mu\nu\alpha\beta}\right)D_\alpha^{ab}\psi_\beta^b+\bar{\phi}^a\partial_\mu D_\mu^{ab}\phi^b+\right.\nonumber\\
&+&\left.gf^{abc}\bar{\phi}^a\partial_\mu\left(c^b\psi^c_\mu\right)\right]\;.\label{gfaction}
\end{eqnarray}

\section{Renormalizability}

The first step in the study of the renormalizability of a theory is to write the Ward identities of the model in a consistent way. Henceforth, we need to introduce some external sources \cite{Piguet:1995er}, in order to control the non-linear nature of the BRST transformations, in the form of BRST doublets, three of them to be precise\footnote{In \cite{Brandhuber:1994uf}, only two source doublets were used. Our third set is needed to control the new symmetry (see \eqref{cnl1}), which is nonlinear.}, namely,
\begin{eqnarray}
s\tau^a_\mu&=&\Omega_\mu^a\;,\;\;\;\;\;\;\;\;s\Omega_\mu^a\;=\;0\;,\nonumber\\
sE^a&=&L^a\;,\;\;\;\;\;\;\;\;\;sL^a\;=\;0\;,\nonumber\\
s\Lambda_{\mu\nu}^a&=&K_{\mu\nu}^a\;,\;\;\;\;sK_{\mu\nu}^a\;=\;0\;.\label{brst3}
\end{eqnarray}
The corresponding quantum number of the external sources are displayed in Table \ref{table2}. The respective external action is given by
\begin{eqnarray}
S_{ext}&=&s\int d^4z\left(\tau_\mu^aD_\mu^{ab}c^b+\frac{g}{2}f^{abc}E^ac^bc^c+gf^{abc}\Lambda^a_{\mu\nu}c^b\bar{\chi}^c_{\mu\nu}\right)\nonumber\\
&=&\int d^4z\left[\Omega_\mu^aD_\mu^{ab}c^b+\frac{g}{2}f^{abc}L^ac^bc^c+gf^{abc}K^a_{\mu\nu}c^b\bar{\chi}^c_{\mu\nu}+\tau^a_\mu\left(D_\mu^{ab}\phi^b+gf^{abc}c^b\psi_\mu^c\right)+\right.\nonumber\\
&+&\left.gf^{abc}E^ac^b\phi^c+gf^{abc}\Lambda^a_{\mu\nu}c^bB^c_{\mu\nu}-gf^{abc}\Lambda^a_{\mu\nu}\phi^b\bar{\chi}^c_{\mu\nu}-\frac{g^2}{2}f^{abc}f^{bde}\Lambda^a_{\mu\nu}\bar{\chi}^c_{\mu\nu}c^dc^e\right]\;.\label{extaction}
\end{eqnarray}
The full action we shall consider is then
\begin{equation}
\Sigma=S_o[A]+S_{gf}+S_{ext}\;.\label{fullaction}
\end{equation}

\begin{table}[h]
\centering
\setlength{\extrarowheight}{.5ex}
\begin{tabular}{cc@{\hspace{-.3em}}cccccc}
\cline{1-1} \cline{3-8}
\multicolumn{1}{|c|}{Source} & & \multicolumn{1}{|c|}{$\tau$} & \multicolumn{1}{c|}{$\Omega$} & \multicolumn{1}{c|}{$E$} & \multicolumn{1}{c|}{$L$} & \multicolumn{1}{c|}{$\Lambda$} & \multicolumn{1}{c|}{$K$}
\\ \cline{1-1} \cline{3-8} 
\\[-1.18em]
\cline{1-1} \cline{3-8}
\multicolumn{1}{|c|}{Dim} & & \multicolumn{1}{|c|}{3} & \multicolumn{1}{c|}{3} & \multicolumn{1}{c|}{4} & \multicolumn{1}{c|}{4} & \multicolumn{1}{c|}{2} & \multicolumn{1}{c|}{2} \\
\multicolumn{1}{|c|}{Ghost n$^o$} & & \multicolumn{1}{|c|}{-2} & \multicolumn{1}{c|}{-1} & \multicolumn{1}{c|}{-3} & \multicolumn{1}{c|}{-2} & \multicolumn{1}{c|}{-1} & \multicolumn{1}{c|}{0}
\\ \cline{1-1} \cline{3-8}
\end{tabular}
\caption{Quantum numbers of the external sources.}
\label{table2}
\end{table}

\subsection{Ward identities}\label{WI}

The action \eqref{fullaction} possesses a rich set of Ward identities which we now list.

\begin{itemize}

\item Slavnov-Taylor identity, which expresses the BRST invariance of the full action \eqref{fullaction}:
\begin{equation}
\mathcal{S}(\Sigma)=0\;,\label{st1}
\end{equation}
where
\begin{eqnarray}
\mathcal{S}(\Sigma)&=&\int d^4z\left[\left(\psi^a_\mu-\frac{\delta\Sigma}{\delta\Omega^a_\mu}\right)\frac{\delta\Sigma}{\delta A^a_\mu}+\frac{\delta\Sigma}{\delta\tau^a_\mu}\frac{\delta\Sigma}{\delta\psi^a_\mu}+\left(\phi^a+\frac{\delta\Sigma}{\delta L^a}\right)\frac{\delta\Sigma}{\delta c^a}+\frac{\delta\Sigma}{\delta E^a}\frac{\delta\Sigma}{\delta\phi^a}+\right.\nonumber\\
&+&\left.b^a\frac{\delta\Sigma}{\delta\bar{c}^a}+\bar{\eta}^a\frac{\delta\Sigma}{\delta\bar{\phi}^a}+B^a_{\mu\nu}\frac{\delta\Sigma}{\delta\bar{\chi}^a_{\mu\nu}}+\Omega^a_\mu\frac{\delta\Sigma}{\delta\tau^a_\mu}+L^a\frac{\delta\Sigma}{\delta E^a}+K^a_{\mu\nu}\frac{\delta\Sigma}{\delta\Lambda^a_{\mu\nu}}\right]\;.\label{st2}
\end{eqnarray}

\item Ordinary Landau gauge fixing and Faddeev-Popov anti-ghost equation:
\begin{eqnarray}
\frac{\delta\Sigma}{\delta b^a}&=&\partial_\mu A_\mu^a\;,\nonumber\\
\frac{\delta\Sigma}{\delta\bar{c}^a}-\partial_\mu\frac{\delta\Sigma}{\delta\Omega^a_\mu}&=&-\partial_\mu\psi_\mu^a\;.\label{Lgf1}
\end{eqnarray}

\item Topological Landau gauge fixing and bosonic anti-ghost equation:
\begin{eqnarray}
\frac{\delta\Sigma}{\delta\bar{\eta}^a}&=&\partial_\mu\psi_\mu^a\;,\nonumber\\
\frac{\delta\Sigma}{\delta\bar{\phi}^a}-\partial_\mu\frac{\delta\Sigma}{\delta\tau^a_\mu}&=&0\;.\label{Lgf2}
\end{eqnarray}

\item Bosonic ghost equation:
\begin{equation}
\mathcal{G}^a_\phi\Sigma=\Delta_\phi^a\;,\label{bg1}
\end{equation}
where
\begin{eqnarray}
\mathcal{G}^a_\phi&=&\int d^4z\left(\frac{\delta}{\delta\phi^a}-gf^{abc}\bar{\phi}^b\frac{\delta}{\delta b^c}\right)\;,\nonumber\\
\Delta_\phi^a&=&gf^{abc}\int d^4z\left(\tau_\mu^bA_\mu^c+E^bc^c+\Lambda_{\mu\nu}^b\bar{\chi}_{\mu\nu}^c\right)\;.\label{bg2}
\end{eqnarray}

\item Ordinary Faddeev-Popov ghost equation:
\begin{equation}
\mathcal{G}^a_1\Sigma=\Delta^a\;,\label{Og1}
\end{equation}
where
\begin{eqnarray}
\mathcal{G}^a_1&=&\int d^4z\left[\frac{\delta}{\delta c^a}+gf^{abc}\left(\bar{c}^b\frac{\delta}{\delta b^c}+\bar{\phi}^b\frac{\delta}{\delta\bar{\eta}^c}+\bar{\chi}^b_{\mu\nu}\frac{\delta}{\delta B^c_{\mu\nu}}+\Lambda^b_{\mu\nu}\frac{\delta}{\delta K^c_{\mu\nu}}\right)\right]\;,\nonumber\\
\Delta^a&=&gf^{abc}\int d^4z\left(E^b\phi^c-\Omega_\mu^bA_\mu^c-\tau_\mu^b\psi_\mu^c-L^bc^c+\Lambda_{\mu\nu}^bB_{\mu\nu}^c-K_{\mu\nu}^b\bar{\chi}_{\mu\nu}^c\right)\;.\label{Og2}
\end{eqnarray}

\item Second Faddeev-Popov ghost equation:
\begin{equation}
\mathcal{G}^a_2\Sigma=\Delta^a\;,\label{Sg1}
\end{equation}
where
\begin{equation}
\mathcal{G}^a_2=\int d^4z\left[\frac{\delta}{\delta c^a}-gf^{abc}\left(\bar{\phi}^b\frac{\delta}{\delta\bar{c}^c}+A^b_\mu\frac{\delta}{\delta\psi^c_\mu}+c^b\frac{\delta}{\delta\phi^c}-\bar{\eta}^b\frac{\delta}{\delta b^c}+E^b\frac{\delta}{\delta L^c}\right)\right]\;.\label{Sg2}
\end{equation}

\item Vector supersymmetry\footnote{Written in the form $\mathcal{W}_\mu = \sum_A \delta_\mu \Phi^A \frac{\delta}{\delta\Phi^A}$, the generators $\delta_\mu$ and the BRST operator satisfy a supersymmetric algebra $\{s, \delta_\mu \}= \partial_\mu$.}:
\begin{equation}
\mathcal{W}_\mu\Sigma=0\;,\label{w1}
\end{equation}
where
\begin{eqnarray}
\mathcal{W}_\mu &=&\int d^4z\left[\partial_\mu A_\nu^a\frac{\delta}{\delta \psi_\nu^a}+\partial_\mu c^a\frac{\delta}{\delta \phi^a}+\partial_\mu\bar{\chi}_{\nu\alpha}^a\frac{\delta}{\delta B_{\nu\alpha}^a}+\partial_\mu\bar{\phi}^a\left(\frac{\delta}{\delta\bar{\eta}^a}+\frac{\delta}{\delta\bar{c}^a}\right)+\right.\nonumber\\
&+&\left.\left(\partial_\mu\bar{c}^a-\partial_\mu\bar{\eta}^a\right)\frac{\delta}{\delta b^a}+\partial_\mu\tau_\nu^a\frac{\delta}{\delta \Omega_\nu^a}+\partial_\mu E^a\frac{\delta}{\delta L^a}+\partial_\mu\Lambda^a_{\nu\alpha}\frac{\delta}{\delta K_{\nu\alpha}^a}\right]\;.\label{w2}
\end{eqnarray}

\item Bosonic non-linear symmetry:
\begin{equation}
\mathcal{T}(\Sigma)=0\;,\label{cnl1}
\end{equation}
where
\begin{equation}
\mathcal{T}(\Sigma)=\int d^4z\left[\frac{\delta\Sigma}{\delta\Omega^a_\mu}\frac{\delta\Sigma}{\delta\psi^a_\mu}-\frac{\delta\Sigma}{\delta L^a}\frac{\delta\Sigma}{\delta\phi^a}-\frac{\delta\Sigma}{\delta K^a_{\mu\nu}}\frac{\delta\Sigma}{\delta B^a_{\mu\nu}}+\left(\bar{c}^a-\bar{\eta}^a\right)\left(\frac{\delta\Sigma}{\delta\bar{c}^a}+\frac{\delta\Sigma}{\delta\bar{\eta}^a}\right)\right]\;.\nonumber\\
\label{cnl2}
\end{equation}

\item Global ghost supersymmetry:
\begin{equation}
\mathcal{G}_3\Sigma=0\;,\label{gss1}
\end{equation}
where
\begin{equation}
\mathcal{G}_3=\int d^4z\left[\bar{\phi}^a\left(\frac{\delta}{\delta\bar{\eta}^a}+\frac{\delta}{\delta\bar{c}^a}\right)-c^a\frac{\delta}{\delta\phi^a}+\tau^a_\mu\frac{\delta}{\delta\Omega^a_\mu}+2E^a\frac{\delta}{\delta L^a}+\Lambda^a_{\mu\nu}\frac{\delta}{\delta K^a_{\mu\nu}}\right]\;.\label{gss2}
\end{equation}

\end{itemize}

We remark that the Faddeev-Popov ghost equations \eqref{Og1} and \eqref{Sg1} can be combined to obtain an exact global supersymmetry,
\begin{equation}
\Delta\mathcal{G}^a\Sigma=0\;,\label{cg1}
\end{equation}
where
\begin{eqnarray}
\Delta\mathcal{G}^a&=&\mathcal{G}_1^a-\mathcal{G}_2^a\;=\;\int d^4z\;f^{abc}\left[\left(\bar{c}^b-\bar{\eta}^b\right)\frac{\delta}{\delta b^c}+\bar{\phi}^b\left(\frac{\delta}{\delta\bar{\eta}^c}+\frac{\delta}{\delta\bar{c}^c}\right)+A_\mu^b\frac{\delta}{\delta\psi_\mu^c}+\right.\nonumber\\
&+&\left.\bar{\chi}^b_{\mu\nu}\frac{\delta}{\delta B_{\mu\nu}^c}+c^b\frac{\delta}{\delta\phi^c}+\Lambda^b_{\mu\nu}\frac{\delta}{\delta K_{\mu\nu}^c}+\tau_\mu^b\frac{\delta}{\delta\Omega_\mu^c}+E^b\frac{\delta}{\delta L^c}\right]\;.\label{cg2}
\end{eqnarray}
We observe the similarity of the equation \eqref{cg1} with the vector supersymmetry \eqref{w1}. It is also worth mentioning that, even though the ghost number of the operator \eqref{cg2} is $-1$, resembling an anti-BRST symmetry, it is not a genuine anti-BRST symmetry. See for instance \cite{Braga:1999ui} for the explicit anti-BRST symmetry in topological gauge theories.

\subsection{Most general counterterm}

In order to construct the most general counterterm consistent with the Ward identities of Sec.~\ref{WI} we add to the classical action \eqref{fullaction} a generic integrated polynomial local in the fields and sources with mass dimension four $\Sigma^c$,
\begin{equation}
\Gamma^{(1)}=\Sigma+\epsilon\Sigma^c\;,\label{ct1}
\end{equation}
where $\epsilon$ is a small perturbative parameter. Obviously, $\Gamma^{(1)}$ is recognized as the quantum action at first order in perturbation theory. Due to the recursive nature of algebraic renormalization theory \cite{Piguet:1995er}, to impose the validity of the Ward identities to $\Gamma^{(1)}$ is equivalent to impose their validity to $\Gamma$ at all orders in perturbation theory. Hence, imposing \eqref{st1}, \eqref{Lgf1}, \eqref{Lgf2}, \eqref{Og1}, \eqref{Sg1}, \eqref{w1}, \eqref{cnl1}, and \eqref{gss1} to $\Gamma^{(1)}$ we find that the most general counterterm that can be added to the classical action must obey
\begin{eqnarray}
\mathcal{S}_\Sigma\Sigma^c&=&0\;,\label{c1}\\
\frac{\delta\Sigma^c}{\delta b^a}&=&0\;,\label{c2}\\
\frac{\delta\Sigma^c}{\delta\bar{c}^a}-\partial_\mu\frac{\delta\Sigma^c}{\delta\Omega^a_\mu}&=&0\;,\label{c3}\\
\frac{\delta\Sigma^c}{\delta\bar{\eta}^a}&=&0\;,\label{c4}\\
\frac{\delta\Sigma^c}{\delta\bar{\phi}^a}-\partial_\mu\frac{\delta\Sigma^c}{\delta\tau^a_\mu}&=&0\;,\label{c5}\\
\mathcal{G}^a_\phi\Sigma^c&=&0\;,\label{c6}\\
\mathcal{G}^a_1\Sigma^c&=&0\;,\label{c7}\\
\mathcal{G}^a_2\Sigma^c&=&0\;,\label{c8}\\
\mathcal{W}_\mu\Sigma^c&=&0\;,\label{c9}\\
\mathcal{T}_{\Sigma}\Sigma^c&=&0\;,\label{c10}\\
\mathcal{G}_3\Sigma^c&=&0\;,\label{c12}
\end{eqnarray}
where $\mathcal{S}_\Sigma$ and $\mathcal{T}_{\Sigma}$ are the linear versions of the operators \eqref{st2} and \eqref{cnl1}, respectively. In fact, the linearized Slavnov-Taylor operator is given by
\begin{eqnarray}
\mathcal{S}_\Sigma&=&\int d^4z\left[\left(\psi^a_\mu-\frac{\delta\Sigma}{\delta\Omega^a_\mu}\right)\frac{\delta}{\delta A^a_\mu}-\frac{\delta\Sigma}{\delta A^a_\mu}\frac{\delta}{\delta\Omega^a_\mu}+\frac{\delta\Sigma}{\delta\tau^a_\mu}\frac{\delta}{\delta\psi^a_\mu}+\left(\Omega^a_\mu+\frac{\delta\Sigma}{\delta\psi^a_\mu}\right)\frac{\delta}{\delta\tau^a_\mu}+\right.\nonumber\\
&+&\left.\left(\phi^a+\frac{\delta\Sigma}{\delta L^a}\right)\frac{\delta}{\delta c^a}+\frac{\delta\Sigma}{\delta c^a}\frac{\delta}{\delta L^a}+\frac{\delta\Sigma}{\delta E^a}\frac{\delta}{\delta\phi^a}+\left(L^a+\frac{\delta\Sigma}{\delta\phi^a}\right)\frac{\delta}{\delta E^a}+\right.\nonumber\\
&+&\left.b^a\frac{\delta}{\delta\bar{c}^a}+\bar{\eta}^a\frac{\delta}{\delta\bar{\phi}^a}+B^a_{\mu\nu}\frac{\delta}{\delta\bar{\chi}^a_{\mu\nu}}+K^a_{\mu\nu}\frac{\delta}{\delta\Lambda^a_{\mu\nu}}\right]\;,\label{st3}
\end{eqnarray}
while the linearized operator $\mathcal{T}_{\Sigma}$ is given by 
\begin{eqnarray}
\mathcal{T}_{\Sigma}&=&\int d^4z\left[\frac{\delta\Sigma}{\delta\Omega^a_\mu}\frac{\delta}{\delta\psi^a_\mu}-\frac{\delta\Sigma}{\delta\psi^a_\mu}\frac{\delta}{\delta\Omega^a_\mu}-\frac{\delta\Sigma}{\delta L^a}\frac{\delta}{\delta\phi^a}-\frac{\delta\Sigma}{\delta\phi^a}\frac{\delta}{\delta L^a}+\frac{\delta\Sigma}{\delta K^a_{\mu\nu}}\frac{\delta}{\delta B^a_{\mu\nu}}+\frac{\delta\Sigma}{\delta B^a_{\mu\nu}}\frac{\delta}{\delta K^a_{\mu\nu}}+\right.\nonumber\\
&+&\left.\left(\bar{c}^a-\bar{\eta}^a\right)\left(\frac{\delta}{\delta\bar{\eta}^a}+\frac{\delta}{\delta\bar{c}^a}\right)\right]\;.\label{fnl3}
\end{eqnarray}

The operator $\mathcal{S}_\Sigma$, being nilpotent, defines a cohomology while the constraint \eqref{c1} represents a cohomology problem for $\Sigma^c$. The cohomology is exactly the same as in \cite{Brandhuber:1994uf}, which is trivial\footnote{The only difference is the extra BRST doublet $(\Lambda,K)$, introduced due to the non-linearity of the bosonic symmetry $\mathcal{T}$, which also belongs to the trivial sector of the cohomology \cite{Piguet:1995er}.}. Hence, the Slavnov-Taylor identity is anomaly-free and the solution of \eqref{c1} is
\begin{equation}
\Sigma^c=\mathcal{S}_\Sigma\Delta^{(-1)}\;,\label{ct2}
\end{equation}
where $\Delta^{(-1)}$ is an integrated local polynomial in the fields and sources and their derivatives bounded by dimension four and with ghost number -1.

Following \cite{Brandhuber:1994uf}, equations \eqref{c1}-\eqref{c9} imply that the counterterm \eqref{ct2} takes the form 
\begin{eqnarray}
\Sigma^c&=&\mathcal{S}_\Sigma \int d^4z \left\{a_1\left[\left(\Omega^{a}_\mu - \partial_\mu \bar{c}^a\right)A^a_\mu + \left(\tau^{a}_\mu-\partial_\mu \bar{\phi}^a\right)\psi^a_\mu\right] + a_2(\tau^{a}_\mu - \partial_\mu \bar{\phi}^a)\partial_\mu c^a+\right.\nonumber\\
&+&\left.a_3\bar{\chi}^{a}_{\mu\nu}\partial_\mu A^a_\nu + a_4 f^{abc} \bar{\chi}^a_{\mu\nu} A^b_\mu A^c_\nu \right\},\label{contra1}
\end{eqnarray}
where $a_1$, $a_2$, $a_3$ and $a_4$ are arbitrary constant coefficients. Now, applying the bosonic symmetry constraint \eqref{c10}, one can straightforwardly show that
\begin{equation}
a_1 = a_2 = 0,
\end{equation}
and that
\begin{equation}
a_4 = \frac{a_3}{2}.
\end{equation}
Hence, the most general local counterterm obeying the symmetry content of the model is reduced to the simple form
\begin{equation}
\Sigma^c = \mathcal{S}_\Sigma \int d^4z\;a\;\bar{\chi}^a_{\mu\nu} F^a_{\mu\nu}\;,\label{ct3}
\end{equation}
where the parameter $a_4$ was renamed as $a$: the only renormalization parameter allowed by the Ward identities of the model. Explicitly, the counterterm \eqref{ct3} reads
\begin{equation}
\Sigma^c = a\int d^4z \,\{B^a_{\mu\nu} F^a_{\mu\nu} - 2\bar{\chi}^a_{\mu\nu} D^{ab}_\mu\psi^b_\nu - g f^{abc} \bar{\chi}^a_{\mu\nu}c^bF^c_{\mu\nu}\}\;.\label{ctfinal}
\end{equation}

As pointed out in \cite{Brandhuber:1994uf}, the choice of Landau gauges forbids the presence of the counterterm $(F^a_{\mu\nu} \pm \widetilde {F}^a_{\mu\nu})^2$, implying that the Yang-Mills term $F^a_{\mu\nu}F^a_{\mu\nu}$ is not produced at the quantum level. This is in agreement with previous one-loop computations carried out in \cite{Birmingham:1991ty,Brooks:1988jm}.

\subsection{Quantum stability}

Once we have at our disposal the most general counterterm consistent with all Ward identities of the model, we must verify if the counterterm can absorb the divergences arising in the evaluation of Feynman graphs. In other words, if the counterterm \eqref{ctfinal} can be consistently absorbed by the classical action \eqref{fullaction} by means of the multiplicative redefinition of the fields, sources and parameters of the model. Therefore, starting from the equation \eqref{ct1}, we must show that $\Gamma^{(1)}$ is of the form $ \Sigma(\Phi_0, \mathcal{J}_0, g_0)$, where  
\begin{eqnarray}
\Phi_0 &=& Z^{1/2}_\Phi \Phi\;; \quad  \Phi_0 = \{A^a_\mu, \psi^a_\mu, c^a, \bar{c}^a, \phi^a, \bar{\phi}^a, b^a, \bar{\eta}^a, \bar{\chi}^a_{\mu\nu}, B^a_{\mu\nu}\}\;,\nonumber\\
\mathcal{J}_0 &=& Z_\mathcal{J}\mathcal{J}\;; \quad \;\;\mathcal{J} = \{\tau^a_\mu, \Omega^a_\mu, E^a, L^a, \Lambda^a_{\mu\nu}, K^a_{\mu\nu}\}\;,\nonumber\\
g_0 &=& Z_g g\;.
\end{eqnarray}
In fact, a direct and straightforward analysis shows that the model is stable. From the gauge fixing action, we obtain
\begin{equation}\label{Z1}
Z_g Z^{1/2}_A = Z^{1/2}_{\bar{c}} Z^{1/2}_c = Z^{1/2}_b Z^{1/2}_A = Z^{1/2}_{\bar{\eta}} Z^{1/2}_\psi = Z^{1/2}_{\bar{c}} Z^{1/2}_\psi = Z^{1/2}_{\bar{\phi}} Z^{1/2}_\phi = Z_g Z^{1/2}_{\bar{\phi}} Z^{1/2}_c  Z^{1/2}_\psi = 1\;,
\end{equation}
and
\begin{equation} \label{Z2}
Z^{1/2}_B Z^{1/2}_A = Z^{1/2}_{\bar{\chi}} Z^{1/2}_c = Z^{1/2}_{\bar{\chi}} Z^{1/2}_\psi =1 + \epsilon a.
\end{equation}
For the source action, we find
\begin{equation}
 Z_\Omega  Z^{1/2}_c =Z_\tau  Z^{1/2}_\phi =  Z_g Z_\tau Z^{1/2}_c  Z^{1/2}_\psi = Z_g Z_L Z_c = Z_g Z_E Z^{1/2}_c Z^{1/2}_\phi  = 1\;,\label{Z3}
\end{equation}
and
\begin{equation}
 Z_g Z_J Z^{1/2}_c Z^{1/2}_{\bar{\chi}} =  Z_g Z_\Lambda Z^{1/2}_\phi Z^{1/2}_{\bar{\chi}} = Z^2_g Z_\Lambda Z_c Z^{1/2}_{\bar{\chi}} = Z_g  Z_\Lambda Z^{1/2}_c Z^{1/2}_B = 1.\label{Z4}
\end{equation}

The results \eqref{Z1}-\eqref{Z4} are self consistent and show that the model is renormalizable to all orders in perturbation theory. 

It is worth mentioning again that the Ward identities \eqref{st1}-\eqref{gss1} hold at all orders with the classical action $\Sigma$ replaced by the 1PI generating functional $\Gamma$. In addition, we would like to emphasize that the result \eqref{ctfinal} is a direct consequence of the absence of anomalies in the Slavnov-Taylor identitiy. The anomalous Slavnov-Taylor identity would give $S_\Sigma \Sigma^c = \bigtriangleup^{(1)}$, being $\bigtriangleup^{(1)}$ a local polynomial with ghost number 1; but it was proven in \cite{Brandhuber:1994uf} that the cohomology of the linearized BRST operator vanishes, therefore, there is no room for anomaly in the Slavnov-Taylor identity, which automatically restricts the most general counterterm of the theory to the trivial sector of the cohomology. As a consequence of this triviality, the cohomology vanishes in any ghost number sector.

\section{Consequences of the Ward identities for the two-point functions}\label{2PFs}

In this Section we provide some strong consequences of the Ward identities in terms of the two-point functions of the theory. Specifically, we compute exact properties\footnote{By exact we mean valid to all orders in perturbation theory. In most cases, this means tree-level exact, \emph{i.e.}, all radiative corrections vanish.} of the propagators and 1PI two-point functions. The conventions and notation here employed can be found in the App.~\ref{AP1}. Needless to say, since the theory is renormalizable to all orders in perturbation theory, the Ward identities are valid for the quantum action $\Gamma$ and not only for the classical one $\Sigma$. 

First of all, we evoke the discrete Faddeev-Popov symmetry (dFPs) to recall that all two-point functions carrying a non-vanishing ghost-number vanish, namely,
\begin{equation}
\Gamma_{(\Phi^A\Phi^B)}(p)=\langle\Phi^A\Phi^B\rangle(p)=0\;\;\forall\;\;g_A+g_B\ne0\;.
\end{equation}
Second, from Lorentz covariance it is easy to infer that we must have, for the (anti)self-dual fields, 
\begin{eqnarray}
\langle b^aB^b_{\mu\nu}\rangle(p)&=&0\;,\label{bB1}\\
\langle c^a\bar{\chi}^b_{\mu\nu}\rangle(p)&=&0\;,\label{cchi1}
\end{eqnarray}
and
\begin{eqnarray}
\Gamma^{ab}_{(b B)\mu\nu}(p)&=&0\;,\label{bB2}\\
\Gamma^{ab}_{(c\bar{\chi})\mu\nu}(p)&=&0\;.\label{cchi2}
\end{eqnarray}

\subsection{1PI two-point functions}

Since the Ward identities are written for the 1PI generating functional, it is easier to start with the 1PI two-point functions. All 1PI two-point functions obtained in this subsection are displayed in Table \ref{table3}.

\subsubsection{Consequences of the Landau gauge fixings}

The ordinary Landau gauge fixing \eqref{Lgf1}, in terms of the quantum action, is given by
\begin{equation}
\frac{\delta\Gamma}{\delta b^a(x)}=\partial_\mu^xA_\mu^a(x)\;,\label{Lgf100a}
\end{equation}
where $\partial_\mu^x$ stands for the spacetime derivative with respect to the coordinates of the point $x_\mu$. In the same way, the topological Landau gauge fixing \eqref{Lgf2} can be written as
\begin{equation}
\frac{\delta\Gamma}{\delta\bar{\eta}^a(x)}=\partial_\mu^x\psi_\mu^a(x)\;.\label{Lgf200a}
\end{equation}

\begin{itemize}
\item \emph{The $bA$ mixed 1PI function.}

To obtain the $bA$ mixed 1PI function, we vary the equation \eqref{Lgf100a} with respect to $A_\nu^b(y)$,
\begin{equation}
\frac{\delta^2\Gamma}{\delta A_\nu^b(y)\delta b^a(x)}=\delta^{ab}\partial_\nu^x\delta(x-y)\;.\label{Lgf100b}
\end{equation}
Hence,
\begin{equation}
\Gamma_{(bA)\nu}^{ab}(x,y)=\delta^{ab}\partial_\nu^x\delta(x-y)\;.\label{Lgf100c}
\end{equation}
Taking the Fourier transform of eq. \eqref{Lgf100c} one obtains
\begin{equation}
\int\frac{d^4p}{(2\pi)^4}\Gamma_{(bA)\mu}^{ab}(p)e^{ip(x-y)}=\int\frac{d^4p}{(2\pi)^4}\delta^{ab}ip_\mu e^{ip(x-y)}\;.\label{Lgf100d}
\end{equation}
Thus,
\begin{equation}
\Gamma_{(bA)\mu}^{ab}(p)=i\delta^{ab}p_\mu\;.\label{bA1}
\end{equation}
The mixed two-point vertex function \eqref{bA1} is tree-level exact, as expected from the relation $Z_bZ_A=1$ in \eqref{Z1}.

\item \emph{The $bb$ 1PI function.}

In the same way, by varying \eqref{Lgf100a} with respect to $b^b(y)$, one trivially finds
\begin{equation}
\Gamma_{(bb)}^{ab}(p)=0\;.\label{bb1}
\end{equation}

\item \emph{The $\bar{\eta}\psi$ mixed 1PI function.}

Now, varying the equation \eqref{Lgf200a} with respect to  $\psi^b_\nu(y)$ and  Fourier transforming the resulting equation, one finds
\begin{equation}
\Gamma_{(\bar{\eta}\psi)\mu}^{ab}(p)=i\delta^{ab}p_\mu\;,\label{etapsi1}
\end{equation}
which is in accordance with the relation $Z_{\bar{\eta}}Z_\psi=1$ in \eqref{Z1}.

\item \emph{The $\bar{\eta} c$ mixed 1PI function.}

And, the variation  of \eqref{Lgf200a} with respect to $c^a(y)$ leads to
\begin{equation}
\Gamma_{(\bar{\eta}c)}^{ab}(p)=0\;.\label{etac1}
\end{equation}

\end{itemize}

\subsubsection{Consequences of the vector supersymmetry}

The vector supersymmetry \eqref{w1}, in terms of the 1PI generating functional, reads\footnote{For simplicity, only the relevant terms are written in \eqref{w3}.}
\begin{eqnarray}
\int d^4z\left[\partial_\gamma A_\kappa^c\frac{\delta\Gamma}{\delta \psi_\kappa^c}+\partial_\gamma c^c\frac{\delta\Gamma}{\delta \phi^c}+\partial_\gamma\bar{\chi}_{\sigma\kappa}^c\frac{\delta\Gamma}{\delta B_{\sigma\kappa}^c}+\partial_\gamma\bar{\phi}^c\left(\frac{\delta\Gamma}{\delta\bar{\eta}^c}+\frac{\delta\Gamma}{\delta\bar{c}^c}\right)\right.&+&\nonumber\\
+\left.\left(\partial_\gamma\bar{c}^c-\partial_\gamma\bar{\eta}^c\right)\frac{\delta\Gamma}{\delta b^c}+\ldots\right]&=&0\;.\label{w3}
\end{eqnarray}

\begin{itemize}

\item \emph{The $BB$ 1PI function.}

Varying \eqref{w3} with respect to $B^b_{\alpha\beta}(y)$ and  $\bar{\chi}^a_{\mu\nu}(x)$ we get
\begin{equation}
\int d^4z\left[\delta^{ac}\delta_{\mu\sigma}\delta_{\nu\kappa}\partial_\gamma^z\delta(z-x)\frac{\delta^2\Gamma}{\delta B_{\alpha\beta}^b(y)\delta B_{\sigma\kappa}^c(z)}+\ldots\right]=0\;.\label{w3a}
\end{equation}
After integration over $z$, a Fourier transformation of \eqref{w3a} yields 
\begin{equation}
p_\gamma\Gamma_{(BB)\mu\nu\alpha\beta}^{ab}(p)=0\;,\label{w3b}
\end{equation}
which, by contraction with $p_\gamma/p^2$, simply reduces to
\begin{equation}
\Gamma_{(BB)\mu\nu\alpha\beta}^{ab}(p)=0\;.\label{BB1}
\end{equation}

\item \emph{The topological ghost and the $BA$ 1PI functions.}

In the same way, by varying with respect to $\bar{\chi}^a_{\alpha\beta}(x)$ and $A^b_\mu(y)$, one finds 
\begin{equation}
-\int d^4z\left[\delta(z-y)\partial_\kappa^z\frac{\delta^2}{\delta\bar{\chi}_{\alpha\beta}^a(x) \delta\psi^b_\mu(z)}+\delta(z-x)\partial_\kappa^z\frac{\delta^2}{\delta A^b_\mu(y)\delta B_{\alpha\beta}^a(z)}+\ldots\right]=0\;.
\end{equation}
Hence,
\begin{equation}
-\partial_\kappa^y\Gamma_{(\bar{\chi}\psi)\alpha\beta\mu}^{ab}(x,y)+ \partial_\kappa^x\Gamma_{(BA)\alpha\beta\mu}^{ab}(x,y)=0\;.
\end{equation}
Fourier transforming this last equation (with attention to the point where the derivative is taken), one obtains
\begin{equation}
\Gamma_{(\bar{\chi}\psi)\alpha\beta\mu}^{ab}(p)=-\Gamma_{(BA)\alpha\beta\mu}^{ab}(p)\;.\label{chipsi1}
\end{equation}
The relation \eqref{chipsi1} is consistent with the relations \eqref{Z2} by means of $Z_BZ_A=Z_{\bar{\chi}}Z_\psi$. Moreover, it is easy to infer from the antisymmetry in $\alpha$ and $\beta$ indices that they should be transverse,
\begin{equation}
\Gamma_{(\bar{\chi}\psi)\alpha\beta\mu}^{ab}(p)=-\Gamma_{(BA)\alpha\beta\mu}^{ab}(p)=X_1(p^2)\epsilon_{\alpha\beta\mu\nu}p_\nu+y(p^2)\left(\delta_{\alpha\mu}p_\beta-\delta_{\beta\mu}p_\alpha\right)\;,\label{chipsi2}
\end{equation}
where $X_1(p^2)$ and $y(p^2)$ are  generic form factors.

\item \emph{The Faddeev-Popov and bosonic ghost 1PI functions.}

Another consequence of the vector supersymmetry concerns the Faddeev-Popov ghost and the bosonic ghost 1PI two-point functions. By varying \eqref{w3} with respect to $c^a(y)$ and $\bar{\phi}^b(x)$, one gets (the proof is very similar to the one displayed in the demonstration of \eqref{chipsi1})
\begin{equation}
\Gamma_{(\bar{\phi}\phi)}^{ab}(p)=\Gamma_{(\bar{c}c)}^{ab}(p)\;.\label{cc1}
\end{equation}
where \eqref{etac1} was used. Expression \eqref{cc1} is in harmony with the relation $Z_{\bar{c}}Z_c=Z_{\bar{\phi}}Z_\phi$ in \eqref{Z1}.

\item \emph{The $\bar{c}\psi$ mixed 1PI function.}

In the same lines of \eqref{chipsi1} and \eqref{cc1}, by varying \eqref{w3} with respect to $\phi^a(x)$ and $\psi^b_\mu(x)$, one can prove that
\begin{eqnarray}
\Gamma_{(\bar{c}\psi)\mu}^{ab}(p)=-\Gamma_{(\bar{\eta}\psi)\mu}^{ab}(p)=-i\delta^{ab}p_\mu\;,\label{cpsi1}
\end{eqnarray}
where \eqref{etapsi1} must be employed. The tree-level exactness \eqref{cpsi1} is in accordance with the relation $Z_{\bar{c}}Z_\psi=Z_{\bar{\eta}}Z_\psi=1$ and the fact that $Z_\psi=Z_c$ and $Z_{\bar{\eta}}=Z_{\bar{c}}$, all coming from the relations \eqref{Z1}. 

\item \emph{The topological gluon 1PI function.}

Now, we consider the topological gluon vacuum polarization $\Gamma_{(AA)\mu\nu}^{ab}(p)$. Remarkably, as can be verified in the App.~\ref{AP2}, it identically vanishes,
\begin{equation}
\Gamma_{(AA)\mu\nu}^{ab}(p)=0\;.\label{AA1}
\end{equation}
\end{itemize}
We will discuss this result in more details in Sec.~\ref{EXTRA}.

\begin{table}[h]
\centering
\setlength{\extrarowheight}{.5ex}
\begin{tabular}{ccc@{\hspace{-.3em}}cccccccccc}
\cline{1-2} \cline{4-13}
\multicolumn{1}{|c|}{$\downarrow\;\Phi^A$} & \multicolumn{1}{c|}{$\Phi^B\rightarrow$} &  & \multicolumn{1}{|c|}{$A^b_\alpha$} & \multicolumn{1}{c|}{$\psi^b_\alpha$} & \multicolumn{1}{c|}{$c^b$} & \multicolumn{1}{c|}{$\phi^b$} & \multicolumn{1}{c|}{$\bar{c}^b$} & \multicolumn{1}{c|}{$b^b$} & \multicolumn{1}{c|}{${\bar{\phi}}^b$} & \multicolumn{1}{c|}{$\bar{\eta}^b$} & \multicolumn{1}{c|}{$\bar{\chi}^b_{\alpha\beta}$} & \multicolumn{1}{c|}{$B^b_{\alpha\beta}$} \\\cline{1-2} \cline{4-13}
\\[-1.18em]
\cline{1-2} \cline{4-13} 
\multicolumn{2}{|c|}{$A^a_\mu$}                                                       & & \multicolumn{1}{|c|}{0}                                             & \multicolumn{1}{c|}{---}                               & \multicolumn{1}{c|}{---}                              & \multicolumn{1}{c|}{---}                        & \multicolumn{1}{c|}{---}         & \multicolumn{1}{c|}{---}   & \multicolumn{1}{c|}{---}            & \multicolumn{1}{c|}{---}            & \multicolumn{1}{c|}{---}                          & \multicolumn{1}{c|}{---}                 \\ \cline{1-2} \cline{4-13}
\multicolumn{2}{|c|}{$\psi^a_\mu$}                                                    & & \multicolumn{1}{|c|}{0}                                             & \multicolumn{1}{c|}{0}                                 & \multicolumn{1}{c|}{---}                              & \multicolumn{1}{c|}{---}                        & \multicolumn{1}{c|}{---}         & \multicolumn{1}{c|}{---}   & \multicolumn{1}{c|}{---}            & \multicolumn{1}{c|}{---}            & \multicolumn{1}{c|}{---}                          & \multicolumn{1}{c|}{---}                 \\ \cline{1-2} \cline{4-13} 
\multicolumn{2}{|c|}{$c^a$}                                                           & & \multicolumn{1}{|c|}{0}                                             & \multicolumn{1}{c|}{0}                                 & \multicolumn{1}{c|}{0}                                & \multicolumn{1}{c|}{---}                        & \multicolumn{1}{c|}{---}         & \multicolumn{1}{c|}{---}   & \multicolumn{1}{c|}{---}            & \multicolumn{1}{c|}{---}            & \multicolumn{1}{c|}{---}                          & \multicolumn{1}{c|}{---}                 \\ \cline{1-2} \cline{4-13} 
\multicolumn{2}{|c|}{$\phi^a$}                                                        & & \multicolumn{1}{|c|}{0}                                             & \multicolumn{1}{c|}{0}                                 & \multicolumn{1}{c|}{0}                                & \multicolumn{1}{c|}{0}                          & \multicolumn{1}{c|}{---}         & \multicolumn{1}{c|}{---}   & \multicolumn{1}{c|}{---}            & \multicolumn{1}{c|}{---}            & \multicolumn{1}{c|}{---}                          & \multicolumn{1}{c|}{---}                 \\ \cline{1-2} \cline{4-13} 
\multicolumn{2}{|c|}{$\bar{c}^a$}                                                     & & \multicolumn{1}{|c|}{0}                                             & \multicolumn{1}{c|}{$-i\delta^{ab}p_\alpha$}           & \multicolumn{1}{c|}{$\Gamma^{ab}_{(\bar{\phi}\phi)}$} & \multicolumn{1}{c|}{0}                          & \multicolumn{1}{c|}{0}           & \multicolumn{1}{c|}{---}   & \multicolumn{1}{c|}{---}            & \multicolumn{1}{c|}{---}            & \multicolumn{1}{c|}{---}                          & \multicolumn{1}{c|}{---}                 \\ \cline{1-2} \cline{4-13} 
\multicolumn{2}{|c|}{$b^a$}                                                           & & \multicolumn{1}{|c|}{$i\delta^{ab}p_\alpha$}                        & \multicolumn{1}{c|}{0}                                 & \multicolumn{1}{c|}{0}                                & \multicolumn{1}{c|}{0}                          & \multicolumn{1}{c|}{0}           & \multicolumn{1}{c|}{0}     & \multicolumn{1}{c|}{---}            & \multicolumn{1}{c|}{---}            & \multicolumn{1}{c|}{---}                          & \multicolumn{1}{c|}{---}                 \\ \cline{1-2} \cline{4-13} 
\multicolumn{2}{|c|}{$\bar{\phi}^a$}                                                  & & \multicolumn{1}{|c|}{0}                                             & \multicolumn{1}{c|}{0}                                 & \multicolumn{1}{c|}{0}                                & \multicolumn{1}{c|}{$\Gamma^{ab}_{(\bar{c}c)}$} & \multicolumn{1}{c|}{0}           & \multicolumn{1}{c|}{0}     & \multicolumn{1}{c|}{0}              & \multicolumn{1}{c|}{---}            & \multicolumn{1}{c|}{---}                          & \multicolumn{1}{c|}{---}                 \\ \cline{1-2} \cline{4-13} 
\multicolumn{2}{|c|}{$\bar{\eta}^a$}                                                  & & \multicolumn{1}{|c|}{0}                                             & \multicolumn{1}{c|}{$i\delta^{ab}p_\alpha$}            & \multicolumn{1}{c|}{0}                                & \multicolumn{1}{c|}{0}                          & \multicolumn{1}{c|}{0}           & \multicolumn{1}{c|}{0}     & \multicolumn{1}{c|}{0}              & \multicolumn{1}{c|}{0}              & \multicolumn{1}{c|}{---}                          & \multicolumn{1}{c|}{---}                 \\ \cline{1-2} \cline{4-13} 
\multicolumn{2}{|c|}{$\bar{\chi}^a_{\mu\nu}$}                                         & & \multicolumn{1}{|c|}{0}                                             & \multicolumn{1}{c|}{$-\Gamma^{ab}_{(BA)\mu\nu\alpha}$} & \multicolumn{1}{c|}{0}                                & \multicolumn{1}{c|}{0}                          & \multicolumn{1}{c|}{0}           & \multicolumn{1}{c|}{0}     & \multicolumn{1}{c|}{0}              & \multicolumn{1}{c|}{0}              & \multicolumn{1}{c|}{0}                            & \multicolumn{1}{c|}{---}                 \\ \cline{1-2} \cline{4-13} 
\multicolumn{2}{|c|}{$B^a_{\mu\nu}$}                                                  & & \multicolumn{1}{|c|}{$-\Gamma^{ab}_{(\bar{\chi}\psi)\mu\nu\alpha}$} & \multicolumn{1}{c|}{0}                                 & \multicolumn{1}{c|}{0}                                & \multicolumn{1}{c|}{0}                          & \multicolumn{1}{c|}{0}           & \multicolumn{1}{c|}{0}     & \multicolumn{1}{c|}{0}              & \multicolumn{1}{c|}{0}              & \multicolumn{1}{c|}{0}                            & \multicolumn{1}{c|}{0}                   \\ \cline{1-2} \cline{4-13} 
\end{tabular}
\caption{Exact results for the two-point vertex functions $\Gamma_{(\Phi\Phi)}^{AB}(p)$. The traces --- are redundancies since the table is (anti-)symmetric by the line-column exchange.}
\label{table3}
\end{table}

\subsection{Propagators}

Now we focus on the connected two-point functions. With this intent, we have to employ the Legendre transformation \eqref{v1} in the Ward identities. All propagators obtained in this subsection are collected in Table \ref{table4}.

\subsubsection{Consequences of the Landau gauge fixings}

The ordinary Landau gauge fixing  equation \eqref{Lgf1}, in terms of the connected Green functional, takes the form
\begin{equation}
-J_{(b)}^a(x)=\partial_\mu^x\frac{\delta W}{\delta J_{(A)\mu}^a(x)}\;,\label{Lgf100e}
\end{equation}
while the topological gauge fixing equation \eqref{Lgf2} turns into
\begin{equation}
J_{(\bar{\eta})}^a(x)=\partial_\mu^x\frac{\delta W}{\delta J_{(\psi)\mu}^a(x)}\;.\label{Lgf200e}
\end{equation}

\begin{itemize}
\item  \emph{The $bA$ mixed propagator.}

Variation of equation \eqref{Lgf100e} with respect to $J_{(b)}^b(y)$ leads to
\begin{equation}
\delta^{ab}\delta(x-y)=\partial_\mu^x\langle A_\mu^a(x)b^b(y)\rangle\;.
\end{equation}
This equation is easily solved in momentum space. Its Fourier transformation leads to
\begin{equation}
\delta^{ab}\int\frac{d^4p}{(2\pi)^4}e^{ip(x-y)}=\partial_\mu^x\int\frac{d^4p}{(2\pi)^4}e^{ip(x-y)}\langle A_\mu^ab^b\rangle(p)\;,
\end{equation}
providing
\begin{equation}
\delta^{ab}=ip_\mu \langle A_\mu^ab^b\rangle(p)\;,
\end{equation}
whose solution is
\begin{equation}
\langle b^aA_\mu^b\rangle(p)=i\delta^{ab}\frac{p_\mu}{p^2}\;.\label{Ab2}
\end{equation}
This is in complete accordance with the relation $Z_bZ_A=1$ in \eqref{Z1}.

\item \emph{The $BA$ mixed propagator.}

The variation of equation \eqref{Lgf100e} with respect to $J_{(B)\alpha\beta}^b(y)$ leads to  the the transversality of  $\langle B^a_{\alpha\beta}A^b_\mu\rangle(p)$, which is evident from the antisymmetry of its indices $\alpha$ and $\beta$. Hence, the $BA$ propagator must be of the form
\begin{equation}
\langle B^a_{\alpha\beta}A^b_\mu\rangle(p)=B_1(p^2)\epsilon_{\alpha\beta\mu\nu}p_\nu+B_2(p^2)\left(\delta_{\alpha\mu}p_\beta-\delta_{\beta\mu}p_\alpha\right)\;,\label{BA2}
\end{equation}
where $B_1(p^2)$ and $B_2(p^2)$ are generic form factors.

\item \emph{The $\bar{\eta}\psi$ mixed propagator.}

Now, varying equation \eqref{Lgf200e} with respect to $J_{(\psi)\nu}^b(y)$ and following the lines in the obtention of \eqref{Ab2}, we get
\begin{equation}
\langle \bar{\eta}^a\psi_\mu^b\rangle(p)=i\delta^{ab}\frac{p_\mu}{p^2}\;.\label{etapsi2}
\end{equation}
The exact result \eqref{etapsi2} is consistent  with  $Z_{\bar{\eta}}Z_\psi=1$ in \eqref{Z1}.

\item \emph{The $\bar{c}\psi$ mixed propagator.}

At last, by varying equation \eqref{Lgf200e} with respect to $J_{(\bar{c})}^b(y)$, a transversality condition is gained (after Fourier transformation),
\begin{equation}
p_\mu\langle\bar{c}^a\psi_\mu^b\rangle(p)=0\;.
\end{equation}
However, from Lorentz covariance, the only possibility is that  $\langle\bar{c}^a\psi_\mu^b\rangle(p)=\delta^{ab}P(p^2)p_\mu$. Thus, inevitably, $P(p^2)=0$, leading to
\begin{equation}
\langle \bar{c}^a\psi_\mu^b\rangle(p)=0\;.\label{cpsi2}
\end{equation}
\end{itemize}

\subsubsection{Consequences of the vector supersymmetry}

In terms of the connected Green functional the vector supersymmetry \eqref{w1} reads
\begin{eqnarray}
& &\int d^4z\left[\partial_\gamma^z\frac{\delta W}{\delta J_{(A)\kappa}^c(z)} J_{(\psi)\kappa}^c(z)-\partial_\gamma^z\frac{\delta W}{\delta J_{(c)}^c(z)}J_{(\phi)}^c(z)-\partial_\gamma^z\frac{\delta W}{\delta J_{(\bar{\chi})\kappa\sigma}^c(z)}J_{(B)\kappa\sigma}^c(z)+\right.\nonumber\\
&+&\left.\partial_\gamma^z\frac{\delta W}{\delta J_{(\bar{\phi})}^c(z)}\left(J_{(\bar{\eta})}^c(z)+J_{(\bar{c})}^c(z)\right)-\partial_\gamma^z\left(\frac{\delta W}{\delta J_{(\bar{c})}^c(z)}-\frac{\delta W}{\delta J_{(\bar{n})}^c(z)}\right)J_{(b)}^c(z)+\ldots\right]=0\;.\nonumber\\
\label{w4}
\end{eqnarray}

\begin{itemize}
\item  \emph{The topological gluon propagator.}

The topological gluon propagator is obtained by varying equation \eqref{w4} with respect to $J_{(A)\mu}^a(x)$ and $J_{(\psi)\nu}^a(y)$,
\begin{equation}
\int d^4z\left[\partial_\gamma^z\frac{\delta^2W}{\delta J_{(A)\mu}^a(x)\delta J_{(A)\kappa}^c(z)} \delta^{bc}\delta_{\nu\kappa}\delta(z-y)+\ldots\right]=0\;.
\end{equation}
Hence, after integration in $z$ and a Fourier transformation, we get
\begin{equation}
p_\gamma\langle A^a_\mu A^b_\nu\rangle(p)=0\;.
\end{equation}
By contraction with $p_\gamma/p^2$, we obtain
\begin{equation}
\langle A^a_\mu A^b_\nu\rangle(p)=0\;.\label{AA2}
\end{equation}
Thus, the topological gluon propagator vanishes just like  the associated vacuum polarization \eqref{AA1}. See  Sec.~\ref{EXTRA} for extra discussions about this issue.

\item  \emph{The Faddeev-Popov and bosonic ghost propagators.}

The relation between the Faddeev-Popov ghost propagator $\bar{c}c$ and the bosonic ghost propagator  $\bar{\phi}\phi$ is obtained by varying equation \eqref{w4} with respect to $J_{(\bar{c})}^a(x)$ and $J_{(\phi)}^b(y)$,
\begin{equation}
\int d^4z\left[-\partial_\gamma^z\frac{\delta^2W}{\delta J_{(\bar{c})}^a(x)\delta J_{(c)}^c(z)}\delta^{cb}\delta(z-y)+\partial_\gamma^z\frac{\delta^2W}{\delta J_{(\phi)}^b(y)\delta J_{(\bar{\phi})}^c(z)}\delta^{ca}\delta(z-x)+\ldots\right]=0\;,
\end{equation}
which reduces to
\begin{equation}
\partial_\gamma^y\langle\bar{c}^a(x)c^b(y)\rangle+\partial_\gamma^x\langle\bar{\phi}^a(x)\phi^b(y)\rangle=0\;.
\end{equation}
Thus, after a Fourier transformation, we get
\begin{equation}
\langle\bar{c}^ac^b\rangle(p)=\langle\bar{\phi}^a\phi^b\rangle(p)\;,\label{cc2}
\end{equation}
which confirms, once again, the relation $Z_{\bar{c}}Z_c=Z_{\bar{\phi}}Z_\phi$ in \eqref{Z1}. We refer to Sec.~\ref{EXTRA} for the proof of the tree-level exactness of the ghost (Faddeev-Popov and bosonic) propagator.

\item \emph{The topological ghost and the mixed $BA$ propagators.}

The topological ghost propagator $\langle \bar{\chi}\psi\rangle$ can be computed by varying \eqref{w4} with respect to $J_{(\bar{\psi})\mu}^a(x)$ and $J_{(B)_{\alpha\beta}}^b(y)$,
\begin{equation}
\int d^4z\left[\partial_\gamma^z\frac{\delta^2W}{\delta J_{(B)\alpha\beta}^b(y)\delta J_{(A)\mu}^a(z)}\delta(z-x)-\partial_\gamma^z\frac{\delta^2W}{\delta J_{(\psi)\mu}^a(x)\delta J_{(\bar{\chi})\alpha\beta}^b(z)}\delta(z-y)+\ldots\right]=0\;,
\end{equation}
which reduces to
\begin{equation}
\partial_\gamma^y\langle\bar{\chi}_{\alpha\beta}^b(y)\psi_\mu^a(x)\rangle-\partial_\gamma^x\langle A_\mu^a(x)B_{\alpha\beta}^b(y)\rangle=0\;.
\end{equation}
After a Fourier transformation, we get
\begin{equation}
\langle\bar{\chi}_{\alpha\beta}^b\psi_\mu^a\rangle(p)=-\langle B_{\alpha\beta}^bA_\mu^a\rangle(p)\;.\label{chipsi3}
\end{equation}
The result \eqref{chipsi3} agrees with \eqref{chipsi1} and with $Z_{\bar{\chi}}Z_\psi=Z_BZ_A$ in \eqref{Z2}.

\item \emph{The $\bar{\eta}c$ mixed propagator.}

Following the same reasoning as before, we vary equation \eqref{w4} with respect to $J_{(c)}^a(x)$ and $J_{(B)\alpha\beta}^b(y)$ and find that
\begin{equation}
\langle\bar{\eta}^ac^a\rangle(p)=\langle \bar{c}^ac^a\rangle(p)\;.\label{etac2}
\end{equation}
This relation is consistent with $Z_{\bar{\eta}}=Z_{\bar{c}}$ in \eqref{Z1}.

\end{itemize}

\begin{table}[h]
\centering
\setlength{\extrarowheight}{.5ex}
\begin{tabular}{ccc@{\hspace{-.3em}}cccccccccc}
\cline{1-2} \cline{4-13}
\multicolumn{1}{|c|}{$\downarrow\;\Phi^A$} & \multicolumn{1}{c|}{$\Phi^B\rightarrow$} &  & \multicolumn{1}{|c|}{$A^b_\alpha$}                                  & \multicolumn{1}{c|}{$\psi^b_\alpha$}                   & \multicolumn{1}{c|}{$c^b$}                            & \multicolumn{1}{c|}{$\phi^b$}                   & \multicolumn{1}{c|}{$\bar{c}^b$} & \multicolumn{1}{c|}{$b^b$} & \multicolumn{1}{c|}{${\bar{\phi}}^b$} & \multicolumn{1}{c|}{$\bar{\eta}^b$} & \multicolumn{1}{c|}{$\bar{\chi}^b_{\alpha\beta}$} & \multicolumn{1}{c|}{$B^b_{\alpha\beta}$} \\\cline{1-2} \cline{4-13}
\\[-1.18em]
\cline{1-2} \cline{4-13} 
\multicolumn{2}{|c|}{$A^a_\mu$}                                                       & & \multicolumn{1}{|c|}{0}                                             & \multicolumn{1}{c|}{---}                               & \multicolumn{1}{c|}{---}                              & \multicolumn{1}{c|}{---}                        & \multicolumn{1}{c|}{---}         & \multicolumn{1}{c|}{---}   & \multicolumn{1}{c|}{---}            & \multicolumn{1}{c|}{---}            & \multicolumn{1}{c|}{---}                          & \multicolumn{1}{c|}{---}                 \\ \cline{1-2} \cline{4-13}
\multicolumn{2}{|c|}{$\psi^a_\mu$}                                                    & & \multicolumn{1}{|c|}{0}                                             & \multicolumn{1}{c|}{0}                                 & \multicolumn{1}{c|}{---}                              & \multicolumn{1}{c|}{---}                        & \multicolumn{1}{c|}{---}         & \multicolumn{1}{c|}{---}   & \multicolumn{1}{c|}{---}            & \multicolumn{1}{c|}{---}            & \multicolumn{1}{c|}{---}                          & \multicolumn{1}{c|}{---}                 \\ \cline{1-2} \cline{4-13} 
\multicolumn{2}{|c|}{$c^a$}                                                           & & \multicolumn{1}{|c|}{0}                                             & \multicolumn{1}{c|}{0}                                 & \multicolumn{1}{c|}{0}                                & \multicolumn{1}{c|}{---}                        & \multicolumn{1}{c|}{---}         & \multicolumn{1}{c|}{---}   & \multicolumn{1}{c|}{---}            & \multicolumn{1}{c|}{---}            & \multicolumn{1}{c|}{---}                          & \multicolumn{1}{c|}{---}                 \\ \cline{1-2} \cline{4-13} 
\multicolumn{2}{|c|}{$\phi^a$}                                                        & & \multicolumn{1}{|c|}{0}                                             & \multicolumn{1}{c|}{0}                                 & \multicolumn{1}{c|}{0}                                & \multicolumn{1}{c|}{0}                          & \multicolumn{1}{c|}{---}         & \multicolumn{1}{c|}{---}   & \multicolumn{1}{c|}{---}            & \multicolumn{1}{c|}{---}            & \multicolumn{1}{c|}{---}                          & \multicolumn{1}{c|}{---}                 \\ \cline{1-2} \cline{4-13} 
\multicolumn{2}{|c|}{$\bar{c}^a$}                                                     & & \multicolumn{1}{|c|}{0}                                             & \multicolumn{1}{c|}{0}           & \multicolumn{1}{c|}{$\left\langle\bar{\phi}^a\phi^b\right\rangle$} & \multicolumn{1}{c|}{0}                          & \multicolumn{1}{c|}{0}           & \multicolumn{1}{c|}{---}   & \multicolumn{1}{c|}{---}            & \multicolumn{1}{c|}{---}            & \multicolumn{1}{c|}{---}                          & \multicolumn{1}{c|}{---}                 \\ \cline{1-2} \cline{4-13} 
\multicolumn{2}{|c|}{$b^a$}                                                           & & \multicolumn{1}{|c|}{$i\delta^{ab}p_\alpha/p^2$}                        & \multicolumn{1}{c|}{0}                                 & \multicolumn{1}{c|}{0}                                & \multicolumn{1}{c|}{0}                          & \multicolumn{1}{c|}{0}           & \multicolumn{1}{c|}{0}     & \multicolumn{1}{c|}{---}            & \multicolumn{1}{c|}{---}            & \multicolumn{1}{c|}{---}                          & \multicolumn{1}{c|}{---}                 \\ \cline{1-2} \cline{4-13} 
\multicolumn{2}{|c|}{$\bar{\phi}^a$}                                                  & & \multicolumn{1}{|c|}{0}                                             & \multicolumn{1}{c|}{0}                                 & \multicolumn{1}{c|}{0}                                & \multicolumn{1}{c|}{$\left\langle\bar{c}^ac^b\right\rangle$} & \multicolumn{1}{c|}{0}           & \multicolumn{1}{c|}{0}     & \multicolumn{1}{c|}{0}              & \multicolumn{1}{c|}{---}            & \multicolumn{1}{c|}{---}                          & \multicolumn{1}{c|}{---}                 \\ \cline{1-2} \cline{4-13} 
\multicolumn{2}{|c|}{$\bar{\eta}^a$}                                                  & & \multicolumn{1}{|c|}{0}                                             & \multicolumn{1}{c|}{$i\delta^{ab}p_\alpha/p^2$}            & \multicolumn{1}{c|}{$\left\langle\bar{c}^ac^b\right\rangle$}                                & \multicolumn{1}{c|}{0}                          & \multicolumn{1}{c|}{0}           & \multicolumn{1}{c|}{0}     & \multicolumn{1}{c|}{0}              & \multicolumn{1}{c|}{0}              & \multicolumn{1}{c|}{---}                          & \multicolumn{1}{c|}{---}                 \\ \cline{1-2} \cline{4-13} 
\multicolumn{2}{|c|}{$\bar{\chi}^a_{\mu\nu}$}                                         & & \multicolumn{1}{|c|}{0}                                             & \multicolumn{1}{c|}{$-\left\langle B^a_{\mu\nu}A^b_{\alpha}\right\rangle$} & \multicolumn{1}{c|}{0}                                & \multicolumn{1}{c|}{0}                          & \multicolumn{1}{c|}{0}           & \multicolumn{1}{c|}{0}     & \multicolumn{1}{c|}{0}              & \multicolumn{1}{c|}{0}              & \multicolumn{1}{c|}{0}                            & \multicolumn{1}{c|}{---}                 \\ \cline{1-2} \cline{4-13} 
\multicolumn{2}{|c|}{$B^a_{\mu\nu}$}                                                  & & \multicolumn{1}{|c|}{$-\left\langle\bar{\chi}^a_{\mu\nu}\psi^b_{\alpha}\right\rangle$} & \multicolumn{1}{c|}{0}                                 & \multicolumn{1}{c|}{0}                                & \multicolumn{1}{c|}{0}                          & \multicolumn{1}{c|}{0}           & \multicolumn{1}{c|}{0}     & \multicolumn{1}{c|}{0}              & \multicolumn{1}{c|}{0}              & \multicolumn{1}{c|}{0}                            & \multicolumn{1}{c|}{0}                   \\ \cline{1-2} \cline{4-13} 
\end{tabular}
\caption{Exact results for the propagators $\langle\Phi^A\Phi^B\rangle(p)$. The traces --- are redundancies since the table is (anti-)symmetric by the line-column exchange.}
\label{table4}
\end{table}

\section{Further considerations}\label{EXTRA}

\subsection{Few words about the topological gluon propagator}

In the previous section, an exact proof of the vanishing of the gluon connected two-point function was worked out. In the present subsection, we compute the tree-level gluon propagator and show that its vanishing is very much related to the particular choice of (Landau-type) gauge we have  employed. For this computation, we introduce two gauge parameters $\alpha$ and $\beta$ through the following quadratic terms:
\begin{equation}
-\frac{\alpha}{2}\int d^4z~b^a b^a\qquad \mathrm{and} \qquad -\frac{\beta}{2}\int d^4z~ B^{a}_{\mu\nu}B^{a}_{\mu\nu}\,,
\label{treelevel1}
\end{equation}
where the choice of signs was done in such a way that these gauge parameters are strictly non-negative. Hence, the terms that contribute to the tree-level topological gluon propagator are given by
\begin{equation}
\tilde{S} = \int d^4z\left[b^a\left(\partial_\mu A^a_\mu - \frac{\alpha}{2}b^a\right)+B^{a}_{\mu\nu}\left(F^{a}_{\mu\nu}\pm \tilde{F}^{a}_{\mu\nu}-\frac{\beta}{2}B^{a}_{\mu\nu}\right)\right]\,.
\label{treelevel2}
\end{equation}
By integrating out the auxiliary fields $(b,B)$, one obtains 
\begin{equation}
\tilde{S} = \int d^4z\left[\frac{(\partial_\mu A^a_\mu)^2}{2\alpha}+\frac{(F^{a}_{\mu\nu}\pm \tilde{F}^{a}_{\mu\nu})^2}{2\beta}\right]\,.
\label{treelevel3}
\end{equation}
Keeping just quadratic terms on $A^a_\mu$ leads to
\begin{equation}
\tilde{S}^{\mathrm{quad}} = - \frac{1}{2\alpha}\int d^4z~A^{a}_\mu\partial_\mu \partial_\nu A^a_\nu - \frac{2}{\beta}\int d^4z\left(A^{a}_\mu \partial^2 A^a_\mu - A^{a}_\mu \partial_\mu \partial_\nu A^{a}_\nu\right)\,,
\label{treelevel4}
\end{equation}
which is expressed in momentum space as
\begin{equation}
\tilde{S}^{\mathrm{quad}} = \frac{1}{2}\int \frac{d^4p}{(2\pi)^4}A^{a}_{\mu}(p)\Delta^{ab}_{\mu\nu}A^{b}_\nu(-p)\,,
\label{treelevel5}
\end{equation}
with
\begin{equation}
\Delta^{ab}_{\mu\nu} = \delta^{ab}\left[\frac{4}{\beta}p^2\delta_{\mu\nu}-\left(\frac{4}{\beta}-\frac{1}{\alpha}\right)p_\mu p_\nu\right]\,.
\label{treelevel6}
\end{equation}
Consequently, the tree-level gluon propagator is 
\begin{equation}
\langle A^{a}_\mu A^{b}_\nu\rangle_0(p) = \delta^{ab}\left[\frac{\beta}{4p^2}\left(\delta_{\mu\nu}-\frac{p_{\mu}p_{\nu}}{p^2}\right)+\frac{\alpha}{p^2}\frac{p_\mu p_\nu}{p^2}\right]\,.
\label{treelevel7}
\end{equation}
The gauge condition we have considered throughout this work corresponds to setting $\alpha=\beta=0$. From eq. \eqref{treelevel7} it is clear that, for such a choice, the gluon propagator vanishes at the tree-level (and this property holds to all orders as proved in the last section). Therefore, this choice is extremely peculiar, since when writing the Feynman rules for this theory, every diagram with gluon lines vanishes. Nonetheless, one can easily see that with the appropriate choice of $\beta=4$, the Yang-Mills term is recovered (see \eqref{treelevel2}). As it is well known, the presence of such term leads to deep relations between topological Yang-Mills theories quantized in a certain class of gauges and supersymmetric gauge theories, see \cite{Fucito:1997xm}. 

\subsection{Exactness of the Faddeev-Popov ghost two-point functions}

In this subsection, we give a proof using Wick theorem that the Faddeev-Popov ghosts two-point function is tree-level exact. For this, we use the property defined by eq. \eqref{cc2}. Hence, let us have a closer look at the $\langle \bar{\phi}^{a}(x) \phi^{b}(y)\rangle$. By definition, 
\begin{equation}
\langle \bar{\phi}^{a}(x) \phi^{b}(y)\rangle = \int \left[\mathcal{D}\Phi\right]\bar{\phi}^{a}(x) \phi^{b}(y)\mathrm{e}^{-S_{gf}}=\int \left[\mathcal{D}\Phi\right]\bar{\phi}^{a}(x) \phi^{b}(y)\mathrm{e}^{-S_{int}}\mathrm{e}^{-S_{quad}}\,,
\label{fpghost1}
\end{equation}
with $\Phi$ a shorthand notation for the complete set of fields of the theory (see App.~\ref{AP1}). The actions $S_{quad}$ and $S_{int}$ stand for the quadratic and interacting parts of $S_{gf}$, respectively. The interacting part of $S_{gf}$ is schematically expressed as
\begin{equation}
S_{int}=\int d^4z\left[BAA+\bar{c}Ac+\bar{\chi}cA+\bar{\chi}cAA+\bar{\chi}A\psi + \bar{\phi}A\phi +\bar{\phi}c\psi\right]\,.
\label{fpghost2}
\end{equation}
Therefore, eq. \eqref{fpghost1} is rewritten as
\begin{eqnarray}
\langle \bar{\phi}^{a}(x) \phi^{b}(y)\rangle &=& \int \left[\mathcal{D}\Phi\right]\bar{\phi}^{a}(x) \phi^{b}(y)~\mathrm{exp}\left(-\int d^4z\left[BAA+\bar{c}Ac+\bar{\chi}cA+\bar{\chi}cAA+\right.\right.\nonumber\\
&+&\left.\left.\bar{\chi}A\psi + \bar{\phi}A\phi +\bar{\phi}c\psi\right]\right)\mathrm{e}^{-S_{quad}}\,.
\label{fpghost3}
\end{eqnarray}
As usual, one can expand the exponential for the interacting part, leading to
\begin{eqnarray}
\langle \bar{\phi}^{a}(x) \phi^{b}(y)\rangle &=& \langle \bar{\phi}^{a}(x) \phi^{b}(y)\rangle_0 - \int d^4z \langle \bar{\phi}^{a}(x) \phi^{b}(y)\left[BAA+\bar{c}Ac+\bar{\chi}cA+\bar{\chi}cAA\right.+\nonumber\\
&+&\left.\bar{\chi}A\psi + \bar{\phi}A\phi +\bar{\phi}c\psi\right]_z\rangle_0+\ldots\,.
\label{fpghost4}
\end{eqnarray}
where $\langle\ldots\rangle_0$ means that the expectation value is taken with respect to the quadratic action. As it is apparent from Table~\ref{table4}, the only non-vanishing two-point function involving $(\bar{\phi},\phi)$ is $\langle \bar{\phi}\phi\rangle$. Therefore, we have to single out Wick contractions of $\phi$ with $\bar{\phi}$. Consequently, the first order correction to \eqref{fpghost1} is 
\begin{eqnarray}
&&\int d^4z \langle \bar{\phi}^{a}(x) \phi^{b}(y)\left[BAA+\bar{c}Ac+\bar{\chi}cA+\bar{\chi}cAA
+\bar{\chi}A\psi + \bar{\phi}A\phi +\bar{\phi}c\psi\right]_z\rangle_0=\nonumber\\
&=& \int d^4z \langle \bar{\phi}^{a}(x) \phi^{b}(y)\left[\bar{\phi}A\phi +\bar{\phi}c\psi\right]_z\rangle_0 = 0\,,
\label{fpghost5}
\end{eqnarray}
where we have kept just terms containing $\phi$ and $\bar{\phi}$ since the contraction with any other fields but those vanishes. Going to higher orders renders the insertion of $\bar{\phi}A\phi$ and $\bar{\phi}c\psi$ on integrated spacetime points. The analysis is divided in the following possibilities:
\begin{itemize}
\item We consider just $\bar{\phi}A\phi$ insertions. In this case, the number of $(\bar{\phi},\phi)$ fields is even and is always possible to contract $(\bar{\phi},\phi)$ in pairs. Nevertheless, for each factor $\bar{\phi}A\phi$ introduced, one also introduces an $A$ field which must be contracted with some other field. In the interacting part, the only non-vanishing correlation function involving $A$ is $\langle BA\rangle$. However, this introduces the term $BAA$ containing two $A$ fields and, at the end, one will have to contract $A$ with some field different from $B$, which vanishes. 

\item We consider just $\bar{\phi}c\psi$ insertions. This leads to a mismatch on the pairing of $(\bar{\phi},\phi)$ fields and gives zero automatically. 

\item We consider mixed insertions of $\bar{\phi}A\phi$ and $\bar{\phi}c\psi$. If the insertions are such that there is an odd number of $(\bar{\phi},\phi)$ fields, then it gives zero. If not, one comes back to the first bullet.
\end{itemize}

The conclusion is that one ends up with the exact tree-level relation,
\begin{equation}
\langle \bar{c}^a (x) c^b (y)\rangle = \langle \bar{\phi}^a (x) \phi^b (y) \rangle = \langle \bar{\phi}^a (x) \phi^b (y) \rangle_0\,.
\label{fpghost6}
\end{equation}
Such an argument can be understood by computing the Feynman rules of the theory and noticing that there is no non-vanishing diagram except for the tree-level one for $\langle \bar{\phi}^a (x) \phi^b (y) \rangle$. It is important to emphasize that this is a consequence of the vanishing of the gluon propagator, a feature of the particular gauge choice used in this paper, as discussed in the previous subsection. 

The explicit form of the tree-level Faddeev-Popov ghost propagator is easily computed from the gauge fixing action \eqref{gfaction}, providing
\begin{equation}
\langle \bar{c}^ac^b\rangle(p)= \langle \bar{\phi}^a\phi^b\rangle(p)=\delta^{ab}\frac{1}{p^2}\;.\label{cc3}
\end{equation}
For completeness, one can compute the 1PI two-point functions $\Gamma_{(\bar{c}c)}^{ab}(p)$ and $\Gamma_{(\bar{\phi}\phi)}^{ab}(p)$ from the identity
\begin{equation}
\sum_C\Gamma_{(\Phi_A\Phi_C)}(p)\langle\Phi_C\Phi_B\rangle(p)=-\delta_{AB}\;.\label{id1}
\end{equation}
Choosing $\Phi_A=\bar{c}^a$ and $\Phi_B=\bar{c}^b$, one can straightforwardly find
\begin{equation}
\Gamma_{(\bar{c}c)}^{ab}(p)=\Gamma_{(\bar{\phi}\phi)}^{ab}(p)=\delta^{ab}p^2\;,\label{cc4}
\end{equation}
where \eqref{cc1} was employed.

\subsection{Exactness of the topological ghost two-point functions}

As for the Faddeev-Popov ghosts, it is possible to prove that the topological ghosts $(\bar{\chi},\psi)$ two-point function is tree-level exact. The proof goes in very strict analogy with the Faddeev-Popov ghosts case and, due to this, we will just mention the main points. To do it, we benefit from the relation \eqref{chipsi2} and compute $\langle B_{\alpha\beta}^b (x) A_\mu^a(y)\rangle$ instead. The only non-vanishing contracting involving the $B$ field is with the gauge field $A$ and vice-versa. Hence, looking at the form of the interaction action \eqref{fpghost2}, one sees that the only insertions allowed are those with $BAA$. Therefore,
\begin{eqnarray}
\langle B_{\alpha\beta}^b (x) A_\mu^a(y)\rangle &=&  \langle B_{\alpha\beta}^b (x) A_\mu^a(y)\rangle_0 - \int d^4z \langle B_{\alpha\beta}^b (x) A_\mu^a(y)(BAA)_z\rangle_0+\nonumber\\
&+&\frac{1}{2!}\int d^4z d^4w \langle B_{\alpha\beta}^b (x) A_\mu^a(y)(BAA)_z (BAA)_w\rangle_0+\ldots\,.
\label{topghost1}
\end{eqnarray}
As is easily seen in eq.~\eqref{topghost1}, the number of $A$ fields due to the insertions is always bigger than the number of $B$ fields. Therefore, the gauge fields will have to be contracted with some other field rather than $B$, resulting in vanishing contributions. Again, this is a consequence of the simplifying properties of the gauge condition we have chosen. For the explicit form of the topological ghost tree-level propagator, we refer to \cite{Brooks:1988jm}.

In the same lines of the previous subsection, it is easy to show that the 1PI two-point functions $\Gamma_{(\bar{\chi}\psi)\alpha\beta\mu}^{ab}$ and $\Gamma_{(BA)\alpha\beta\mu}^{ab}$ are also tree-level exact.  The proof follows by setting $\Phi_A=\bar{\chi}^a_{\alpha\beta}$ and $\Phi_A=\bar{\chi}^b_{\mu\nu}$ in \eqref{id1} and employing the propagators derived in \cite{Brooks:1988jm}.

Henceforth, all two-point functions of the present model are tree-level exact.

\section{Conclusions}

In this work we have studied the renormalizability of topological gauge theories \cite{Donaldson:1983wm,Witten:1988ze,Birmingham:1991ty,Baulieu:1988xs} in the light of the algebraic renormalization technique \cite{Piguet:1995er}. It was shown that the the most general counterterm, given by expression \eqref{ctfinal},  has only one independent renormalization parameter. This is a novel result compared with the standard literature \cite{Brandhuber:1994uf}, whose authors found four independent renormalization parameters. The main reason of such reduction in the number of independent renormalizations of topological gauge theories is a new symmetry, the bosonic non-linear symmetry described in \eqref{cnl1}.

The (anti-)self-dual Landau gauges choice \eqref{gf1} provides a remarkable rich set of Ward identities, namely, \eqref{st1}, \eqref{Lgf1}, \eqref{Lgf2}, \eqref{Og1}, \eqref{Sg1}, \eqref{w1}, \eqref{cnl1} and \eqref{gss1}. Making use of these identities, we were able to show that most (connected and vertex) two-point functions are tree-level exact (see tables \ref{table3} and \ref{table4}). The first exceptions are the Faddeev-Popov and the bosonic ghost two-point functions, which are  equal to each other (see \eqref{cc1} and \eqref{cc2}). The second exceptions are the topological ghost and the mixed $BA$ two-point functions, which are also equal to each other (modulo a minus sign) as characterized by \eqref{chipsi1} and \eqref{chipsi2}. Nevertheless, we were able to show by an explicit perturbative expansion that all exceptions are actually tree-level exact as well.

The gauge choice of (anti-)self-dual Landau gauges has another remarkable consequence: the fact that the two-point functions of the topological gluon vanish to all orders in perturbation theory (see expressions \eqref{AA1} and \eqref{AA2}), including the tree-level. This property is, perhaps, the main reason of the tree-level exactness of all two-point functions because a vanishing gauge propagator would eliminate most of the Feynman diagrams. Essentially, this exceptional feature is due to a combined effort from the Slavnov-Taylor identity \eqref{st1} and the vector supersymmetry \eqref{w1}, the latter being an exclusive feature of the (anti-)self-dual Landau gauges.

It could be interesting to explore the three- and four-point functions of topological gauge theories in the (anti-)self-dual Landau gauges. In this way, one could establish, for instance, the running behavior of the coupling parameter. Moreover, the link between the present model and a (non-Wess-Zumino type) supersymmetric $N=2$ gauge theory can also be investigated. Another interesting aspect to analyze are the possible consequences of the exact topological properties in the corresponding supersymmetric theory. Finally, the Gribov problem in supersymmetric theories can also be studied under the light of the topological sector. All of these aspects are currently under investigation.

\section*{Acknowledgements}

ADP wishes to express his gratitude to Silvio P.~Sorella for fruitful discussions. GS is thankful for the warm hospitality at CECs, where part of this work was developed. The Conselho Nacional de Desenvolvimento Cient\'ifico e Tecnol\'ogico (CNPq - Brazil) and the Coordena\c{c}\~ao de Aperfei\c{c}oamento de Pessoal de N\'ivel Superior (CAPES) are acknowledged for financial support.

\appendix

\section{Conventions for Green functions generators}\label{AP1}

In this section we employ the conventions of Euclidean QFT as in \cite{Piguet:1995er}. Let us write the most relevant relations that we will employ. The Green functional is defined as
\begin{equation}
Z[J]=N\int D\Phi e^{-\Sigma-\int d^4z J^A\Phi^A}\;,\label{z1}
\end{equation}
where $N=1/Z[0]$ is the usual normalization, $\Phi^A$ stands for all fields, $J^A$ are Schwinger sources introduced for each field and $A$ is a multiple index ranging all fields. The functional measure is then $D\Phi=\prod_Ad\Phi^A$. The connected Green functional $W[J]$ is defined as
\begin{equation}
e^{-W[J]}=Z[J]\;.\label{w1ap}
\end{equation}
Hence, the quantum action (vertex functional) is given by
\begin{equation}
\Gamma[\Phi]=W[J]-\int d^4z J^A\Phi^A\bigg|_{\Phi^A=\frac{\delta W}{\delta J^A}}\;,\label{v1}
\end{equation}
whose inverse reads
\begin{equation}
W[J]=\Gamma[\Phi]+\int d^4z J^A\Phi^A\bigg|_{J^A=(-1)^{(g_A+1)}\frac{\delta\Gamma}{\delta\Phi^A}}\;,\label{w2ap}
\end{equation}
where $g_A$ stands for the statistics of the field $\Phi^A$ (+1 for fermions and 0 for bosons). And, as usual,
\begin{eqnarray}
\frac{\delta W}{\delta J^A}\Bigg|_{J^A=0}=\frac{\delta\Gamma}{\delta\Phi^A}\Bigg|_{\Phi^A=0}=0\;.\label{vev1}
\end{eqnarray}

The connected two-point functions will be denoted by
\begin{equation}
\langle\Phi^A(x)\Phi^B(y)\rangle=-\frac{\delta^2 W}{\delta J^B(y)\delta J^A(x)}\Bigg|_{J=0}\;.\label{c2pf}
\end{equation}
In momentum space, we have,
\begin{equation}
\langle\Phi^A(x)\Phi^B(y)\rangle=\int\frac{d^4p}{(2\pi)^4}e^{ip(x-y)}\langle\Phi^A\Phi^B\rangle(p)\;.\label{fourier1}
\end{equation}
For the amputated two-point functions we define
\begin{equation}
\Gamma_{\Phi\Phi}^{AB}(x,y)=\frac{\delta^2\Gamma}{\delta\Phi^B(y)\delta\Phi^A(x)}\Bigg|_{\Phi=0}\;,\label{a2pf}
\end{equation}
and the corresponding Fourier transform reads
\begin{equation}
\Gamma_{\Phi\Phi}^{AB}(x,y)=\int\frac{d^4p}{(2\pi)^4}e^{ip(x-y)}\Gamma_{\Phi\Phi}^{AB}(p)\;.\label{fourier2}
\end{equation}

\section{Proof of $\Gamma_{(AA)\mu\nu}^{ab}(p)=0$}\label{AP2}

To proof the  exact result \eqref{AA1}, we consider the Slavnov-Taylor identity \eqref{st1} for the vertex functional $\Gamma$,
\begin{equation}
\mathcal{S}(\Gamma)=\int d^4z\left[\left(\psi^c_\alpha(z)-\frac{\delta\Gamma}{\delta\Omega^c_\alpha(z)}\right)\frac{\delta\Gamma}{\delta A^c_\alpha(z)}+\ldots\right]\;.\label{st1a}
\end{equation}
Varying \eqref{st1a} \emph{w.r.t.} $\psi^a_\mu(x)$ and $A^b_\nu(y)$ we get
\begin{equation}
\int d^4z\left[\left(\delta^{ca}\delta_{\alpha\mu}\delta(z-x)-\frac{\delta^2\Gamma}{\psi^a_\mu(x)\delta\Omega^c_\alpha(z)}\right)\frac{\delta^2\Gamma}{\delta A^b_\nu(y)\delta A^c_\alpha(z)}+\ldots\right]=0\;,\label{st1b}
\end{equation}
which simplifies to
\begin{equation}
\frac{\delta^2\Gamma}{\delta A^b_\nu(y)\delta A^a_\mu(x)}-\int d^4z\left[\frac{\delta^2\Gamma}{\psi^a_\mu(x)\delta\Omega^c_\alpha(z)}\frac{\delta^2\Gamma}{\delta A^b_\nu(y)\delta A^c_\alpha(z)}+\ldots\right]=0\;.\label{st1c}
\end{equation}
At vanishing sources and fields \eqref{st1c} yields
\begin{equation}
\Gamma_{(AA)\mu\nu}^{ab}(x,y)-\int d^4z\left[\frac{\delta^2\Gamma}{\psi^a_\mu(x)\delta\Omega^c_\alpha(z)}\frac{\delta^2\Gamma}{\delta A^b_\nu(y)\delta A^c_\alpha(z)}\right]^{J^A=0}_{\Phi^A=0}=0\;.\label{st1d}
\end{equation}
Now, to show that the second term in \eqref{st1d} vanishes we develop
\begin{eqnarray}
\frac{\delta^2\Gamma}{\psi^a_\mu(x)\delta\Omega^c_\alpha(z)}&=&\sum_A\int d^4w\frac{\delta^2 W}{\delta J_{(\Phi)}^A(w)\delta\Omega^c_\alpha(z)}\frac{\delta J_{(\Phi)}^A(w)}{\delta\psi^a_\mu(x)}\nonumber\\
&=&\sum_A(-1)^{g_A+1}\int d^4w\frac{\delta^2 W}{\delta J_{(\Phi)}^A(w)\delta\Omega^c_\alpha(z)}\frac{\delta^2 \Gamma}{\delta\psi^a_\mu(x)\delta\Phi^A(w)}\;.
\end{eqnarray}
Evoking the dFPs, the only fields $\Phi^A$ that may generate non-vanishing two-point functions are the  fields with ghost number $-1$. Hence
\begin{eqnarray}
\frac{\delta^2\Gamma}{\psi^a_\mu(x)\delta\Omega^c_\alpha(z)}&=&\int d^4w\left[\frac{\delta^2 W}{\delta J_{(\bar{c})}^d(w)\delta\Omega^c_\alpha(z)}\frac{\delta^2 \Gamma}{\delta\psi^a_\mu(x)\delta\bar{c}^d(w)}+\frac{\delta^2 W}{\delta J_{(\bar{\eta})}^d(w)\delta\Omega^c_\alpha(z)}\frac{\delta^2 \Gamma}{\delta\psi^a_\mu(x)\delta\bar{\eta}^d(w)}+\right.\nonumber\\
&+&\left.\frac{\delta^2 W}{\delta J_{(\bar{\chi})\sigma\gamma}^d(w)\delta\Omega^c_\alpha(z)}\frac{\delta^2 \Gamma}{\delta\psi^a_\mu(x)\delta\bar{\chi}_{\sigma\gamma}^d(w)}\right]\;.
\end{eqnarray}
At vanishing sources and fields, this last expression reads
\begin{eqnarray}
\frac{\delta^2\Gamma}{\psi^a_\mu(x)\delta\Omega^c_\alpha(z)}&=&\int d^4w\left[\langle D^{ce}_\alpha c^e(z)\bar{c}^d(w)\rangle\Gamma_{(\bar{c}\psi)\mu}^{da}(w,x)+\langle D^{ce}_\alpha c^e(z)\bar{\eta}^d(w)\rangle\Gamma_{(\bar{\eta}\psi)\mu}^{da}(w,x)+\right.\nonumber\\
&+&\left.\langle D^{ce}_\alpha c^e(z)\bar{\chi}_{\sigma\gamma}^d(w)\rangle\Gamma_{(\bar{\chi}\psi)\sigma\gamma\mu}^{da}(w,x)\right]\;.
\end{eqnarray}
It is easy to see, from the BRST transformations \eqref{brst1} and \eqref{brst2}, that the above composite propagators can be written as (omitting the spacetime dependence and indices)
\begin{eqnarray}
\langle Dc\bar{c}\rangle&=&-\langle s(A\bar{c})\rangle+\langle \psi\bar{c}\rangle+\langle Ab\rangle\;=\;\langle \psi\bar{c}\rangle+\langle Ab\rangle\;,\nonumber\\
\langle Dc\bar{\eta}\rangle&=&-\langle s(A\bar{\eta})\rangle+\langle \psi\bar{\eta}\rangle\;=\;\langle \psi\bar{\eta}\rangle\;,\nonumber\\
\langle Dc\bar{\chi}\rangle&=&-\langle s(A\bar{\chi})\rangle+\langle \psi\bar{\chi}\rangle+\langle AB\rangle\;=\langle \psi\bar{\chi}\rangle+\langle AB\rangle\;,
\end{eqnarray}
where the known fact that the expectation value of BRST exact quantities are zero was used (see, for instance, \cite{Piguet:1995er,Becchi:1975nq,Kugo:1979gm} and references therein).  Moreover, due to \eqref{cpsi2} and \eqref{chipsi3}, we get
\begin{eqnarray}
\langle Dc\bar{c}\rangle&=&\langle Ab\rangle\;,\nonumber\\
\langle Dc\bar{\eta}\rangle&=&-\langle s(A\bar{\eta})\rangle+\langle \psi\bar{\eta}\rangle\;=\;\langle \psi\bar{\eta}\rangle\;,\nonumber\\
\langle Dc\bar{\chi}\rangle&=&0\;,
\end{eqnarray}
Hence,
\begin{eqnarray}
\frac{\delta^2\Gamma}{\psi^a_\mu(x)\delta\Omega^c_\alpha(z)}&=&\int d^4w\left[\langle A^c_\alpha(z)b^d(w)\rangle\Gamma_{(\bar{c}\psi)\mu}^{da}(w,x)+\langle \psi^c_\alpha(z)\bar{\eta}^d(w)\rangle\Gamma_{(\bar{\eta}\psi)\mu}^{da}(w,x)\right]\nonumber\\
&=&\int d^4w\left[\langle A^c_\alpha(z)b^d(w)\rangle-\langle \psi^c_\alpha(z)\bar{\eta}^d(w)\rangle\right]\Gamma_{(\bar{c}\psi)\mu}^{da}(w,x)\nonumber\\
&=&0\;,
\end{eqnarray}
where, in the second line, we used the fact that $\Gamma_{(\bar{c}\psi)\mu}^{da}(w,x)=-\Gamma_{(\bar{\eta}\psi)\mu}^{da}(w,x)$ (see \eqref{etapsi1} and \eqref{cpsi1}). In the third line, the relations \eqref{Ab2} and \eqref{etapsi2} were employed. Therefore, we finally achieve
\begin{equation}
\Gamma_{(AA)\mu\nu}^{ab}(x,y)=0\;,\label{st1e}
\end{equation}
as we wanted to show.

\addcontentsline{toc}{section}{References}
\bibliographystyle{utphys2}
\bibliography{library}

\end{document}